\documentstyle[12pt,epsf]{article}

\renewcommand{\bar}[1]{\overline{#1}}

\newcommand{\bra}[1]{\langle  {#1}  \vert }
\newcommand{\ket}[1]{\vert {#1} \rangle }

\begin{document}
\input epsf

\begin{flushright}
SLAC--PUB--8353\\
January 25, 2000
\end{flushright}
\bigskip\bigskip

\thispagestyle{empty}
\flushbottom

\centerline{{\Large\bf 
CORE and the Haldane Conjecture}
\footnote{Work supported in part by Department of Energy contracts
DE--AC03--76SF00515 and DE--AC02--76ER03069}}
\vspace{22pt}
\centerline{\bf Marvin Weinstein}
\vspace{8pt}
\centerline{\it Stanford Linear Accelerator Center}
\centerline{\it Stanford University, Stanford, California 94309}
\vfill
\begin{center}
ABSTRACT
\end{center}

\vfill
The Contractor Renormalization group formalism (CORE) is a real-space
renormalization group method which is the Hamiltonian
analogue of the Wilson exact renormalization group equations.
In an earlier paper\cite{QGAF} I showed that the Contractor Renormalization group (CORE)
method could be used to map a theory of free quarks, and quarks interacting
with gluons, into a generalized frustrated Heisenberg antiferromagnet (HAF)
and proposed using CORE methods to study these theories.
Since generalizations of HAF's exhibit all sorts of subtle behavior
which, from a continuum point of view, are related to topological properties
of the theory, it is important to know that CORE can be used to extract this
physics. 
In this paper I show that despite the folklore which asserts that all
real-space renormalization group schemes are necessarily inaccurate,
simple Contractor Renormalization group (CORE) computations can give highly
accurate results even if one only keeps a small number of states per block
and a few terms in the cluster expansion.  In addition I argue that even
very simple CORE computations give a much better qualitative understanding
of the physics than naive renormalization group methods.  In particular
I show that the simplest CORE computation yields a first principles
understanding of how the famous Haldane conjecture works for the case of
the spin-$1/2$ and spin-$1$ HAF.

\vfill

\begin{center}
Submitted to Physical Review D.
\end{center}
\vfill  
\newpage

\section{Introduction}

The Contractor Renormalization group (CORE) formalism is a Hamiltonian
analogue of the Wilson exact renormalization group equations for systems
defined by a path integral.   Although it is a real-space renormalization
group method it differs from earlier {\it naive\/} real-space renormalization group
methods\cite{rsrg}, or more accurate methods such as the density-matrix
renormalization group approach of S.~R.~White\cite{densitymatrix}, in that
it is in principle exact, is amenable to evaluation by a convergent
non-perturbative approximation procedure and finite range results are
easily improved by a simple extrapolation technique. In addition, the great
flexibility one has in the choice of truncation procedure allows one to apply CORE
to problems in ways which are impossible with other methods.  In
particular, CORE can be used to produce a non-trivial renormalization group
analysis of a lattice gauge-theories truncated to totally gauge-invariant
block states, something which is impossible in the either naive real-space or
the density matrix renormalization group approach.

In an earlier paper\cite{QGAF} I showed that CORE can be used to map a
theory of free quarks, and quarks interacting with gluons, into a generalized
frustrated Heisenberg antiferromagnet (HAF) and proposed using the same CORE
methods to study these theories.  Since generalizations of HAF's exhibit all
sorts of subtle behavior which from a continuum point of view, are related
to topological properties of the theory, it is important to know that CORE
can be used to extract this physics.  Moreover, since the really interesting
cases are Hamiltonian theories in 3-spatial dimensions, it is important that
CORE be able to produce qualitatively and quantitatively correct pictures
of the low energy physics of these theories with truncation schemes that
keep only a few states per block and only a few terms in the cluster expansion.  
The purpose of this paper is to show that, unlike the original naive real-space
renormalization group methods, relatively simple CORE computations based
upon keeping a small number of states per block and only a few terms in the
finite range cluster expansion give accurate results which can be
{\it systematically\/} improved.  It is the fact that one can achieve
reasonable accuracy keeping only a few states per block which makes it
possible to apply CORE to Hamiltonian theories defined on two and
three dimensional lattices.  

A detailed discussion of the application of CORE to the Hamiltonian
version of the 1+1-dimensional Ising model, which was presented in an earlier
paper\cite{COREpaper}, showed that one could achieve highly
accurate results for the groundstate energy density, magnetization
and mass-gap with a scheme which kept only two states per three-site block
and only up to range-3 terms in the cluster expansion (which means that
the biggest problem one has to deal with is a nine-site sublattice). 
The same paper also presented, among other things, a brief discussion
of the method as applied to the spin-1/2 HAF.  Since the purpose
of that discussion was to use the example to explain certain features of
the CORE method I didn't include any remarks about how the method
compares to the naive renormalization group scheme or how to obtain
higher accuracy results.  These issues will be addressed in this paper before
moving on to the issue of what is different about the spin-1 case and how it
relates to Haldane's conjecture\cite{Haldane}.  My purpose in discussing the spin-1 HAF,
is to show that, {\it unlike the naive renormalization
group approach\/}, a simple 4-state range-2 CORE computation for the spin-1 HAF,
is good enough to provide a straightfoward understanding of the physics.
This simple calculation shows that the physics of the spin-1
model is intimately related to the structure of a more general
theory with Hamiltonian $ H = \sum_i [\vec{s}(i)\cdot \vec{s}(i+1) - \beta\,
(\vec{s}(i)\cdot \vec{s}(i+1))^2 ] $ which has a valence bond ground state
when $\beta = -1/3 $.  In a naive renormalization of the same theory
the term $ - \beta\, (\vec{s}(i)\cdot \vec{s}(i+1))^2 ] $
will not appear and thus, the naive renormalization group approach will not
see any difference between the spin-1/2 theory and the spin-1 theories.

\section{The Basic Problem}

The generic real-space renormalization group (RSRG) method consists of three steps.
First one divides the lattice into blocks, each of which contains a finite number
of sites.  Next one restricts the full Hamiltonian to just those terms which
relate to a single block, diagonalizes it, selects a finite subset of its
lowest energy eigenstates and then uses them to generate a subspace of the full
Hilbert space which we will refer to as the set of {\it retained states\/}.
Finally, one constructs a new {\it renormalized\/} Hamiltonian which acts
with the space of retained states and which has the same 
low-energy physics as the full theory.  Of course, the details of how to choose
blocks, how to construct the appropriate block Hamiltonian and how to
construct the renormalized Hamiltonian acting on the space of retained states
differs from method to method.

\subsection{The Naive Real-Space Renormalization Group}

The naive real-space renormalization group method implements the RSRG procedure
in a straightforward manner and is basically a version of
the Rayleigh-Ritz variational calculation familiar from elementary quantum mechanics.
In order to understand the motivation behind the method and its limitations
consider the example of a simple spin-1/2 Heisenberg anti-ferromagnet with the
Hamiltonian 
\begin{equation}
   H = \sum_{j = -\infty}^{\infty} \vec{s}(j) \cdot \vec{s}(j+1) .			
\end{equation}
Begin by dividing the lattice into disjoint blocks each containing three
sites, $B_j = \{ 3j, 3j+1, 3j+2 \}$ and define the block Hamiltonian
\begin{equation}
	H_{B_j} = \vec{s}(3j)\cdot\vec{s}(3j+1) + \vec{s}(3j+1)\cdot \vec{s}(3j+2)
\end{equation}
Next diagonalize the block Hamiltonian and keep its two lowest lying states.
This is a simple exercise since the block Hamiltonian can be rewritten
in terms of the total block spin operator
$\vec{S}_{\rm TOT}(3j,3j+1,3j+2) = \vec{s}(3j) + \vec{s}(3j+1) + \vec{s}(3j+2)$ and the
operator $\vec{S}_{\rm TOT}(3j,3j+2) = \vec{s}(3j) + \vec{s}(3j+2)$, as
\begin{eqnarray}
\label{Naivethreesite}
	H_{B_j} &=& \vec{s}(3j)\cdot\vec{s}(3j+1) + \vec{s}(3j+1)\cdot \vec{s}(3j+2) \\
	&=& \vec{s}(3j+1)\cdot\left( \vec{s}(3j) + \vec{s}(3j+2) \right) \\
	&=&  \frac{1}{2} \left ( S^2_{\rm TOT}(3j,3j+1,3j+2) - S^2_{\rm TOT}(3j,3j+2) - \frac{3}{4} 
			\right)
\end{eqnarray}
The Hilbert space for the three-site problem is a product of three spin-1/2 states
and since $H_{B_j}$ is rotationally invariant its eigenstates decompose into
the direct sum of a spin-3/2 multiplet and two spin-1/2 multiplets.
From Eq.~\ref{Naivethreesite} we see that the lowest lying multiplet is the spin-$1/2$
multiplet obtained by coupling the spin on site $3j+1$ to the spin-1 multiplet made
out of the product of the spins on sites $3j$ and $3j+2$.  Thus, keeping the two
lowest lying states amounts to keeping the lowest lying spin-1/2 multiplet.
We then use these states to generate the subspace of {\it retained
states\/}.

The intuition behind the final step in the naive real-space renormalization
group method, constructing the renormalized Hamiltonian, is based upon
the observation that the gaps between the one block multiplets are fairly large.
One guesses that a reasonable variational
wavefunction for the true groundstate of the system can be constructed
within the space of retained states; i.e., the set generated by taking all
possible tensor products of the two lowest lying spin-1/2 states per block.
A variational calculation based upon this assumption 
says that solving for the best variational state is equivalent to diagonalizing the
Hamiltonian obtained by computing all matrix elements of the original Hamiltonian
in the space of retained states.
To be specific, consider the three-site blocking scheme proposed above.  Let us
denote by $W_j$ the lowest lying spin-1/2 multiplet for block $B_j$ and define
the space of retained states as
\begin{equation}
	W = \bigotimes_j W_j ,
\end{equation}   
Then, if we let $P_W$ denote the projection operator onto the space of retained states
the renormalized Hamiltonian is
\begin{equation}
	H^{\rm ren} = P_W H P_W .
\end{equation}
To explicitly compute $H^{\rm ren}$ it is convenient to rewrite the full
Hamiltonian as a sum of two terms; i.e.
\begin{equation}
\label{termsinsum}
	H = \sum_j H_{B_j} + \sum_j H_{B_j B_{j+1}} \\
\end{equation}
where $H_{B_j}$ is the block Hamiltonian defined in Eq.~\ref{Naivethreesite}
and $H_{B_j B_{j+1}}$ is the block-block coupling term
\begin{equation}
	H_{B_j B_{j+1}} = \vec{s}(3j+2)\cdot\vec{s}(3(j+1)) .
\end{equation}
Since the space of retained states is constructed of tensor products of eigenstates
of the $H_{B_j}$ which all have the eigenvalue $-1$, it is clear that
\begin{equation}
	P_W H_{B_j} P_W = - {\bf 1}_j ,
\end{equation}
and so the first sum of terms in Eq.~\ref{termsinsum} gives a contribution of
$-1$ for each three-site block $B_j$, or in other words, a contribution
of $-1/3 $ to the energy density of the groundstate.

The block-block interaction term is a sum of terms, each of which
touches two adjacent blocks, so to compute $P_W H_{B_j B_{j+1}} P_W $ we only need
to compute the matrix elements of $\vec{s}(3j+1)$ and $\vec{s}(3(j+1))$
between the states in the lowest lying spin-1/2 multiplet of the three-site problem.
This is easily done and the result is that the truncated operators
on the first and last sites of a three-site block are proportional to 
the spin-1/2 generators with a proportionality factor of $2/3$; i.e.,
\begin{equation}
	P_{W_j} \vec{s}(3j+1) P_{W_j} = P_{W_j} \vec{s}(3j) P_{W_j}
	= \frac{2}{3}\vec{s}(j) ,
\end{equation}
where $\vec{s}(j)$ now stands for the usual spin operators acting on the
spin-1/2 representation associated with each site of the new lattice.
Combining these facts we obtain a renormalized Hamiltonian
\begin{equation}
\label{hren}
	H^{\rm ren} = \sum_j (-{\bf 1}_j) + \frac{4}{9} \vec{s}(j)\cdot\vec{s}(j+1)
\end{equation}
where $H^{\rm ren}$ is to be thought of as acting on a the states of a thinner
lattice (one with one-third as many sites) with a spin-1/2 degree of freedom
associated with each site, $j$. Note that I have chosen to include the energy
density as a sum of single-site operators which have a coefficient of $-1$ and
which happen to be the single-site identity operator.

From Eq.~\ref{hren} we see that the Hamiltonian reproduces itself up to an
additive $c$-number and a multiplicative factor of $4/9$.  It follows immediately that
if we repeatedly apply the same naive renormalization group transformation
group, then after $n$ steps the renormalized Hamiltonian
will have the same form; i.e., a $c$-number term which gives the vacuum energy
and an interaction term which is multiplied by a factor $(4/9)^n$
\begin{equation}
	H^{\rm ren}_n = C_n \sum_j {\bf 1}_j + (4/9)^n \sum_j \vec{s}(j)\cdot\vec{s}(j+1)
\end{equation}
where the coefficients $C_n$ satisfy the recursion relation
\begin{equation}
	C_{n+1} = 3 C_n - (4/9)^n
\end{equation}
The first term on the right hand side of this equation comes from the fact that the
term proportional to the unit matrix contributes $3 C_n$ to the energy of every state
in the three-site Hamiltonian and the second term is just the fact that the lowest lying
spin-1/2 multiplet for the three-site problem has energy $-1$ times the scale factor
of the $\vec{s}\cdot\vec{s}$ term.  To extract the groundstate
energy density we observer that after $n$-steps each site on the new lattice has
is equivalent to $3^n$ sites on the old lattice, thus the energy density is
\begin{equation}
	{\cal E} = \lim_{n \rightarrow \infty} {\cal E}_n = \lim_{n \rightarrow \infty} \frac{C_n}{3^n}
\end{equation}
where
\begin{equation}
\label{eseries}
	{\cal E}_{n+1} = \frac{C_{n+1}}{3^{n+1}} =  {\cal E}_n - \frac{1}{3} (4/27)^n
\end{equation}
and ${\cal E}_1 = -1/3$.
Clearly, one can derive a recursion relation for
the ${\cal E}_n$'s and sum the resulting geometric series to get the groundstate energy
density
\begin{equation}
	{\cal E} = \frac{-1}{3(1-4/27)} = -0.3913 ;
\end{equation}
which is to be compared to the exact answer of $-.4431$.  This corresponds to a fractional
error of $12 \%$.  Furthermore, from the fact that the coefficient in front of the
interaction term $\vec{s}(j)\cdot\vec{s}(j+1)$ tends to zero as $n \rightarrow
\infty$ we see that the theory has to be massless.

Although the computation of the groundstate energy density is only good to $12\%$,
(which is not as good as the Anderson spin-wave computation which is accurate to
about $2\%$) it is very simple and one might hope that it can be easily improved upon.
Unfortunately, this is not as easy as it sounds.

An obvious way to try and improve the calculation is to work with larger blocks
and keep a larger number of states per block so as to get a larger space of
retained states and the possibility of a better variational wavefunction.
Appealing as this sounds, the brute force approach of keeping more
of the lowest lying eigenstates doesn't provide a rapid improvement of the
results obtained in the simplest two state approach.
The problem is that when the block is larger the wavefunctions
of the lowest lying states develop nodes at the walls of the block and therefore
the block-block recoupling terms come out smaller than they should be in the
renormalization group step.  If one is going to work with larger
blocks and keep more states one has to be clever about choosing the states to keep.
This is what is done in the density matrix renormalization group approach.
The principle shortcomings of the density matrix approach is that in general, in
order to achieve high accuracy, one has to keep a large number of states per block which
means: first, that the method is purely numerical in character and one loses
contact with the original structure of the Hamiltonian; second, that the method
is really best suited to Hamiltonian theories on a one-dimensional spatial lattice
since the number of states per block which must be kept to guarantee the correct
recoupling across the boundary of the block in higher dimensions grows quickly
and the problem becomes computationally difficult;  third, for the case of a
lattice gauge-theory, one cannot adopt a truncation
scheme which keeps only locally gauge-invariant states, as the density matrix
method will intrinsically require keeping states in which
flux leaves through the boundaries of a block.  This inability to work with locally
color-singlet states makes using the density matrix method unwieldy for extracting
the low energy physics of a theory like lattice QCD.  

CORE takes a different approach to getting improved results.  It is based upon
a formula which, given a truncation scheme for selecting the space of retained
states, maps the original Hamiltonian to one
which acts only on the space of retained states and this new Hamiltonian
is guaranteed to have the same low energy physics as the original theory.
Although CORE is based upon a scheme which, like the naive renormalization
group approach, keeps only a small number of states per block, a CORE transformation
generates new operators.  Thus, one trades in the information carried by the extra
states for extra operators in the {\it renormalized\/} Hamiltonian.  The advantage
of the CORE approach is that, as we will see, the number of extra operators which
must be kept is much smaller than the number of extra states needed for a density
matrix renormalization group calculation.  Since the CORE method preserves
the basic structure of the original theory the semi-analytic nature of the
resulting renormalization group flow reveals what is happening in a more
transparent manner.  Moreover, as was discussed in Ref.\cite{QGAF}, CORE allows
one to study a theory such as lattice QCD by defining the space of retained states
to be that generated by taking tensor products of local color singlet
states.

\subsection{CORE -- The Basic Algorithm} 

CORE has two parts.  The first is a theorem which defines
the Hamiltonian analog of Wilson's exact renormalization group
transformation; the second is a set of approximation procedures which render
{\it nonperturbative calculation of the renormalized Hamiltonian\/}
doable.  A detailed review of the general method can be found in 
Ref.~\cite{QGAF} and a detailed presentation of the CORE
formalism can be found in Ref.~\cite{COREpaper}.  In this section I 
limit myself to a review of the basic concepts for the special case
of a general Heisenberg antiferromagnet.

As in the case of the naive renormalization group, CORE defines the space
of retained states as the image of a projection operator, $P$, acting
on the original space, ${\cal H}$; i.e., ${\cal H}_{\rm ret} = P
{\cal H}$.  In what follows, for both the spin-$1/2$ and spin-$1$ case,
this set of retained states will
be defined by diagonalizing the Hamiltonian restricted to
either a two or three-site block and defining $P$ as the operator which projects
onto the subspace spanned by a small number of its lowest energy eigenstates.  

The formula relating the original Hamiltonian, $H$, to the {\it renormalized
Hamiltonian\/} having the same low energy physics is
 \begin{equation}
        H^{\rm ren} = \lim_{t\rightarrow \infty}
            [[T(t)^2]]^{-1/2}\, [[T(t) H T(t)]] \, [[T(t)^2]]^{-1/2}   ,
\label{basicform}
\end{equation}
where $T(t) = e^{-t H}$ and where $[[O]]= P O P$ for
any operator $O$ which acts on ${\cal H}$.   It is worth noting
that the $t=0$ version of Eq.~\ref{basicform} is just the definition
of the naive renormalization group transformation. 

While it is not generally possible to evaluate Eq.~\ref{basicform}
exactly, it is possible to nonperturbatively approximate the
infinite lattice version of $H^{\rm ren}$ to any desired 
accuracy.  This is because $H^{\rm ren}$, as defined in
Eq.~\ref{basicform}, is an extensive operator and has the general form
 \begin{equation}
        H^{\rm ren} = \sum_j \sum_{r=1}^{\infty} h^{\rm conn}(j,r) 
\label{hcluster}
\end{equation}
where each term, $h^{\rm conn}(j,r)$, stands for a set of {\it range-$r$
connected\/} operators based at site $j$, all of which can be
evaluated to high accuracy using finite size lattices.
Typically it isn't necessary to calculate all
the terms in $H^{\rm ren}$.  Often one can obtain highly accurate results, or
qualitatively correct results, by approximating $H^{\rm ren}$ by its range-2 or range-3
terms.

In general the range-1 connected term in the renormalized Hamiltonian
is defined to be the matrix obtained by evaluating the $j^{\rm th}$
block Hamiltonian in the set of retained eigenstates,
\begin{equation}
\label{honeconn}
   h^{\rm conn}(j,1) = [[ H_{\rm block}(j) ]] .
\end{equation}
The range-2 part of the renormalized Hamiltonian 
is evaluated as follows: first, restrict the full Hamiltonian to
two adjacent (i.e., connected) blocks
and define the two-block retained states as tensor products of the
single block retained states; next, use these states to define a
projection operator and evaluate Eq.~\ref{basicform}, where $H=H(j,j+1)$ is the
Hamiltonian restricted to blocks $B_j$ and $B_{j+1}$ to obtain
\begin{equation}
\label{twoblocklimit}
H_{2-block}(j,j+1) = \lim_{t \rightarrow \infty}
     [[T(t)^2]]^{-1/2}\, [[T(t) H T(t)]] \, [[T(t)^2]]^{-1/2}  ;
\end{equation}
finally, construct the {\it connected range-2\/}  contribution to the
renormalized Hamiltonian by subtracting the two ways
of embedding the one-block computation into the connected two-block
computation as follows,
\begin{equation}
h^{\rm conn}(j,2) = H_{2-block}(j,j+1) - h^{\rm conn}(j,1) -h^{\rm conn}(j+1,1) .
\end{equation}

It might appear to be difficult to take the $t \rightarrow \infty$
limit of Eq.~\ref{twoblocklimit}, however it is easy to show that this limit can be
evaluated as a product of the form
\begin{equation}
H_{2-block}(j,j+1) = R\,H_{\rm diag} \,R^{\dag}
\end{equation}
where $R$ is an orthogonal transformation and $H_{\rm diag}$ is a diagonal matrix.
$H_{\rm diag}$ is constructed by expanding the image under $R$ of each of the tensor
product states in a complete set of eigenstates of the two-block problem and
putting the energy of the lowest lying eigenstate appearing in the expansion
of each rotated state on the diagonal.
$R$ is constructed to guarantee that for each rotated state,
the lowest energy eigenstate of the two-block problem which appears in its
expansion in a complete set of eigenstates is distinct from that
appearing in the expansion of the other rotated
states.  As we will see in a moment, given the symmetries of the problem,
constructing $R$ is straightforward for both the spin-$1/2$ and spin-$1$ HAF.

The generic range-$r$ connected contribution is obtained by
evaluating 
\begin{equation}
\label{rblocklimit}
H_{r-block}(j,j+1,\ldots,j+r-1) = \lim_{t \rightarrow \infty}
     [[T(t)^2]]^{-1/2}\, [[T(t) H T(t)]] \, [[T(t)^2]]^{-1/2}  ;
\end{equation}
for the Hamiltonian restricted to a set of $r$-adjacent blocks.
Finally, the connected range-$r$ contribution to the renormalized Hamiltonian is
then defined as
\begin{eqnarray}
h^{\rm conn}(j,r) &=& H_{r-block}(j,j+1,\ldots,j+r-1) -
\sum_{m=0}^{1} h^{\rm conn}(j+m,r-1) \nonumber\\
&\ldots& - \sum_{m=0}^{p} h^{\rm conn}(j+m,r-p) 
 \ldots - \sum_{m=0}^{r-1}h^{\rm conn}(j+m,1)  .
\end{eqnarray}

\section{Generalized Heisenberg Antiferromagnets}

The general CORE method is extremely flexible since one has a great deal
of freedom in choosing how to truncate the space of states.  Once one commits
to a given truncation algorithm, however, everything is specified
and it only remains to choose how many terms one will compute in the cluster
expansion.  Given these two somewhat independent choices it is interesting to
explore the way in which changing the truncation algorithm and changing the
range of the cluster expansion affects the accuracy of the results obtained.
The next section discusses this issue for the
case of the spin-1/2 Heisenberg antiferromagnet.  In order to explore the
rate of convergence of the cluster expansion I will first discuss 
the extreme case of a single state truncation algorithm computed to range-$6$ in the
cluster expansion and then I discuss the simplest two-state truncation algorithm
computed to range-$4$ in the cluster expansion.  In addition I will discuss
the use of type two Pad\'e approximants to extrapolate the resulting series for
the groundstate energy density in the single state and two state situation.

\subsection{CORE and the Spin-$1/2$ HAF: One State Truncation}

As in the discussion of the naive renormalization group
algorithm for the HAF we begin with the spin-1/2 Hamiltonian 
\begin{equation}
	H = \sum_{j = -\infty}^{\infty} \vec{s}_j \cdot \vec{s}_{j+1}			
\end{equation}
but, this time we consider various blocking and
truncation algorithms.  Before diving in to the computation it is worth explaining why
we didn't consider a two-site blocking procedure in our discussion of the naive
renormalization group method.  The reason becomes obvious if we rewrite the two-site Hamiltonian,
as 
\begin{eqnarray}
\label{Htwosites}	
	H_{\rm block} &=& \vec{s}_1 \cdot \vec{s}_2 =
	\frac{1}{2} \left(\vec{s}_1 + \vec{s}_2 \right)^2 - \frac{3}{4} , \nonumber \\
	&=& \frac{1}{2} S^{2}_{\rm TOT}(1,2) -\frac{3}{4} ,
\end{eqnarray}
where the notation $S^2_{\rm TOT}(1,2)$ is used to represent the total
spin operator for sites $1$ and $2$.  This shows that $H_{\rm block}$ is 
proportional to $S^2_{\rm TOT}$ minus a constant and so the four eigenstates
of the two-site Hilbert space fall into one spin-$0$
representation of energy $E_0 = -3/4$ and one spin-$1$ representation
with energy $E_1 = 1/4$, which means that the spin-$0$ state has the lowest energy.
From this it follows that any algorithm based
upon keeping a subset of the lowest lying eigenstates of $H_{\rm block}$
requires either that we keep the single spin-$0$ state, or that we                                                                      
keep all four eigenstates of $H_{\rm block}$.  Obviously the first choice,
truncating to one state per block, produces a renormalized Hamiltonian
which is a one-by-one matrix, which allows us to only compute
the energy density of the groundstate.  Moreover, if we keep this single state per
block then, in the naive renormalization group computation, the matrix elements
of the operators $\vec{s}(j)$ will all be zero
and the renormalization group computation will immediately terminate.
Thus, we obtain an estimate for the groundstate
energy density equal to  $-3/8$.  This is, of course, terrible.
The other choice, keeping all of the states per two-site block, clearly isn't a
truncation.
  
CORE differs from the naive renormalization group prescription in that
even a single state truncation procedure leads to a
non-trivial formula for the groundstate energy density which can be
systematically approximated using the cluster expansion.
As an example, once again consider the spin-$1/2$ HAF and a truncation
algorithm based upon keeping the lowest lying
spin-$0$ eigenstate of the two-site Hamiltonian.  In this case the spaces $W_j$
are all one-dimensional and therefore $W$ is too.  Thus, the renormalized
Hamiltonian is a single number which is the groundstate energy density
if the product over all of the single block spin-$0$ states has a
non-vanishing overlap with the true groundstate.

The cluster expansion for the groundstate energy density in this single-state
truncation is particularly simple.  We begin by evaluating
Eq.~\ref{basicform} for the two-site block which gives,
of course, the energy of the spin-$0$ state;  i.e.,
\begin{equation}
	\epsilon^{\rm conn}_1 = h^{\rm conn}(j,1) = [[H_{\rm block}(j)]] = -\frac{3}{4}
\end{equation}   
To obtain the range-$2$ term in the cluster expansion we solve the two-block
(or four-site) problem and verify that the tensor product of the two single-block
spin-$0$ states has a non-vanishing overlap with the two-block groundstate.  If
this is true, then the general formula can be written as
\begin{eqnarray}
  \epsilon^{\rm conn}_2 &=& E_0^2 - h^{\rm conn}(j,1) - h^{\rm conn}(j+1,1) 
	\nonumber\\
		&=& E_0^2 - 2\epsilon^{\rm conn} =  E_0^2 - \frac{3}{2} 
\end{eqnarray}
where $E_0^2$ is the groundstate energy of the four-site block.
Similarly, the other terms we will compute are given by
\begin{eqnarray}
	\epsilon^{rm conn}_3 &=& E_0^3 - 2\epsilon^{\rm conn}_2 - 3\epsilon^{\rm conn}_1
	\nonumber\\
\epsilon^{rm conn}_4 &=& E_0^4 - 2\epsilon^{\rm conn}_3 - 3\epsilon^{\rm conn}_2
		-4\epsilon^{\rm conn}_1 \nonumber \\
\epsilon^{rm conn}_5 &=& E_0^5 - 2\epsilon^{\rm conn}_4 - 3\epsilon^{\rm conn}_3
- 4\epsilon^{\rm conn}_2 - 5\epsilon^{\rm conn}_1 \nonumber\\			
\epsilon^{rm conn}_6 &=& E_0^6 - 2\epsilon^{\rm conn}_5 - 3\epsilon^{\rm conn}_4
- 4\epsilon^{\rm conn}_3 - 5\epsilon^{\rm conn}_2 - 6\epsilon^{\rm conn}_1				
\end{eqnarray}
and the range-$r$ approximation to the groundstate energy density is given by
\begin{equation}
	{\cal E}_r = \sum_{m = 1}^r \epsilon^{\rm conn}_m .
\end{equation}
The values for ${\cal E}_r, r=1,\ldots,6$ are shown in the second column of
Table~\ref{spinhalftable} where one sees that the first six terms in the
cluster expansion produce an estimate for the groundstate energy density which
is good to a part in $10^{-3}$.  This shows that the cluster expansion
converges remarkably rapidly.  In fact, if one compares the value obtained at
range-$6$ to Lanczos calculations\cite{spinhalflanczos} done for the same system
on very large lattices, we see that in the range-$6$ error of a part in $10^{-3}$
corresponds to the error obtained in the $28$-site Lanczos calculation.
The authors in Ref.~\cite{spinhalflanczos} use extrapolation methods to obtain
a more accurate answer from these results.  Clearly the finite range cluster
expansion can also be extrapolated as a function of $1/r$ .  A simple
and powerful way to do this is to use Pad\'e approximants.  To be specific,
we fit the sequence ${\cal E}_r$ to a rational polynomial of the general form
\begin{equation}
  {\cal E}_{[n/m]} = \frac{\alpha_0 + \alpha_2 / r^2 + \alpha_3 / r^3 + \ldots + \alpha_{M+1} / r^{M+1}}{1 + \beta_2 / r^2
	+ \beta_3 / r^3 + \ldots + \beta_{N+1} / r^{N+1} }
\end{equation}
(Note the absence of a term proportional to $1/r$ in either the numerator or denominator.
This is because the cluster expansion removes this term.)
Column three in Table.~\ref{spinhalftable} gives the values of $M$ and $N$ used 
to construct an approximating polynomial and column four give the value of $\alpha_0$,
which corresponds to taking the limit $1/r \rightarrow 0$.  As is evident from the table
the error obtained by extrapolating the series obtained from the first six terms
in the cluster expansion is $6\times 10^{-6}$.  This is not as good as that
obtained by extrapolating the first 14 terms in Ref.\cite{spinhalflanczos},
which is a part in $10^{-7}$, but it isn't bad for a computation which only goes out
to a twelve-site lattice instead of a $28$-site lattice.  It is worth pointing
out that the computation shown in Table.~\ref{spinhalftable} was done by
brute force using the new numerical capabilities of Maple6.  The entire computation
took twenty minutes on a PC equipped a 450 Mhz Pentium3 and 512Meg of ram.
A similar result for the groundstate of the spin-$1$ HAF is shown in
Table.~\ref{spinonetable}.  Once again we see that the range-$4$ cluster expansion
gives a value which converges to within $1.8\%$ of the answer obtained from a
sixteen-site Lanczos calculation\cite{moreo}.  The first few Pad\'e approximants
which can be formed from this series improves the accuracy of the result
to $0.3 \%$.

It is worth pointing out that there is no three-site analog of
formula which follows from doing a single state truncation for a two-site block.
The reason for this, as we saw in the discussion of the naive renormalization
group, is that the lowest lying states of a three-site block are a spin-$1/2$ multiplet.
If one keeps only one state in the spin-$1/2$ subspace then taking the tensor
product of this over $n$-blocks produces a totally symmetric spin state
which would has spin $n/2$.  Since the lowest lying states for even $n$ have
spin-$0$ it follows that the retained states obtained in this way won't have
an overlap with the groundstate (or in fact any low lying state) and therefore
the basic CORE formula won't construct a Hamiltonian which reproduces the low
energy physics of the theory.  Fortunately, for the three-site blocking scheme
the prescription that one should keep the lowest lying states implies one
should keep the entire spin-$1/2$ multiplet.  This produces a non-trivial
renormalization group transformation which does work, as I shall show in the
next section.  In any event, these results make it clear that
improving even the simplest CORE results by computing more terms
in the cluster expansion works well.

\subsection{CORE and the Spin-$1/2$ HAF: Two State Truncation}

Working with three-site blocks, as we saw in the discussion of the naive
renormalization group, is as simple as working with
two-site blocks. As we noted, the three-site Hamiltonian has the form
\begin{eqnarray}
\label{Hthreesites}
	H_{\rm block} &=& \vec{s}_1 \cdot \vec{s}_2 + \vec{s}_2 \cdot \vec{s}_3 \\
	&=& \vec{s}_2 \cdot \left( \vec{s}_1 + \vec{s}_3 \right) \\
	&=&  \frac{1}{2} \left ( S^2_{\rm TOT}(1,2,3) - S^2_{\rm TOT}(1,3) - \frac{3}{4} 
			\right) ,
\end{eqnarray}
and as in the case of the naive renormalization group we truncate
the three-site Hilbert space to the lowest lying spin-$1/2$ multiplet.  

If we label the two spin-$1/2$ states which we keep in block $B_j$ as
$\ket{\uparrow_j}$ and $\ket{\downarrow_j}$, then the projection operator
is
\begin{eqnarray}
\label{threesiteproj}
	P_j &=& \ket{\uparrow_j} \bra{\uparrow_j} +
		\ket{\downarrow_j} \ket{\downarrow_j} \nonumber \\
	P &=& \prod_j P_j 
\end{eqnarray}
By definition the connected range-1 Hamiltonian is $P_j\,H_{\rm block}(j)\, P_j$
which, because the two retained states are degenerate, is simply a multiple
of the identity matrix; i.e.,
\begin{equation}
	h^{\rm conn}(j,1) = - {\mathbf 1}  .
\end{equation}
and so, to this range, the renormalized Hamiltonian is
\begin{equation}
	H^{\rm ren} = \sum_j h^{\rm conn}(j,1) = -V_{\rm thin}\, {\mathbf 1} ;
\end{equation}
i.e., every state in the space of retained states is an eigenstate of the
renormalized Hamiltonian with eigenvalue $-V_{\rm thin}$, where $V_{\rm thin}$
is the volume of the thinned lattice.  Note that $V_{\rm thin} = V /3$ and so the
contribution to the energy density of the original theory is $-1/3$.
Clearly, since all retained states are eigenstates of the range-1 part of
the renormalized Hamiltonian, this term plays no role in the dynamics of the
renormalized theory.  To get a nontrivial renormalized Hamiltonian it is
necessary to calculate $h^{\rm conn}(j,2)$.

The first step in computing $h^{\rm conn}(j,2)$ is to expand the
retained states for the two-block problem in terms of the exact
eigenstates of the two-block Hamiltonian.
A brute force way to do this is to exactly diagonalize the
full two-block, or six-site, Hamiltonian, find its eigenvalues and
eigenstates and then carry out the expansion.  This is not an intelligent
use of computing resources.  Since the spin-$1/2$ HAF has so much
symmetry, one can achieve the desired goal
with less work.

The three-site truncation procedure is based upon keeping the two
states of the lowest lying spin-$1/2$ representation of $SU(2)$ for each
three-site block.  Thus, if we are working with blocks $\{B_j, B_{j+1}\}$,
then the four-dimensional space of retained states is spanned by the
four tensor product states 
\begin{equation}
\ket{\uparrow_j}\ket{\uparrow_{j+1}},\, \ket{\uparrow_j}\ket{\downarrow_{j+1}},\,
\ket{\downarrow_j}\ket{\uparrow_{j+1}},\, \ket{\downarrow_j}\ket{\downarrow_{j+1}}\, .
\end{equation}
As stated earlier, to find the matrix $R$ it is necessary to find a set of
orthonormal combinations of these states which contract onto unique eigenstates
of the six-site problem.  While in general this requires expanding the tensor
product states in terms of eigenstates of the six-site problem,
the symmetries of this problem make finding $R$ an exercise in group-theory
because the six-site Hamiltonian has the same $SU(2)$ symmetry
of the full problem and its eigenstates also fall into irreducible
representations of $SU(2)$.

The argument goes as follows. The space of retained states is generated
from a tensor product of two spin-$1/2$ representations and it can
be uniquely decomposed into a direct sum of one spin-$0$ and one spin-$1$ representation.
Furthermore, the three spin-$1$ states can be uniquely identified by their total $S_z$
eigenvalues, $1, 0, -1$.  The linear combinations corresponding to these
$\ket{S,S_z}$ eigenstates are
\begin{eqnarray}
\label{uniquestates}
\ket{0,\phantom{-}0} &=-& \frac{1}{\sqrt{2}}\,\left( \ket{\uparrow_j} \ket{\downarrow_{j+1}} -
				\ket{\downarrow_j} \ket{\uparrow_{j+1}} \right) \nonumber \\
\ket{1,\phantom{-}1} &= & \ket{\uparrow_j} \ket{\uparrow_{j+1}} \nonumber \\
\ket{1,\phantom{-}0} &=& \frac{1}{\sqrt{2}}\,\left( \ket{\uparrow_j}\ket{\downarrow_{j+1}} +
				\ket{\downarrow_j} \ket{\uparrow_{j+1}} \right) \nonumber \\
\ket{1,-1} &= & \ket{\downarrow_j} \ket{\downarrow_{j+1}} \nonumber
\end{eqnarray}
Since $SU(2)$ is an exact symmetry of the six-site problem only
eigenstates of $H_{\rm 6-sites}$ having the same $S$ and $S_z$ can appear
in the expansion of each one of these states; thus it follows directly
from Eq.~\ref{uniquestates} that all one need to find $h^{\rm conn}(j,2)$
is to find the energy of the lowest lying spin-$0$ and lowest lying spin-$1$
multiplet for $H_{6-sites}$.  This observation, coupled with the
fact that the spin-$0$ states is odd under left-right interchange, whereas
the spin-$1$ state is even, reduces the general problem of diagonalizing
a $64 \times 64$-matrix to that of diagonalizing a couple of
$3 \times 3$-matrices.  As the states in the spin-$1$ multiplet are degenerate
the result of this calculation is an $H_{\rm diag}$ of the form
\begin{equation}
	H_{\rm diag} = \left(
	\begin{array}{c c c c}
	\epsilon_0 & 0 & 0 & 0 \\
	0 & \epsilon_1 & 0 & 0 \\
	0 & 0 & \epsilon_1 & 0 \\
        0 & 0 & 0 & \epsilon_1
	\end{array} \right)
\end{equation}
Using Eq.~\ref{uniquestates} it is simple to compute $R^\dag\,H_{\rm diag}\, R$
acting on the original tensor product states.  Fortunately, one can 
avoid doing even this amount of work.  Due to the $SU(2)$ symmetry
of the theory $R^\dag\,H_{\rm diag}\, R$ must have the form
\begin{equation}
 R^\dag\,H_{\rm diag}\,R = 
	\alpha_0 {\mathbf 1} + \alpha_1 \vec{s}_j \cdot \vec{s}_{j+1}
\end{equation}
To relate $\alpha_0$ and $\alpha_1$ to $\epsilon_0$ and $\epsilon_1$ use the usual
trick and rewrite $R^\dag\,H_{\rm diag}\,R$ as
\begin{eqnarray}
\label{gentwositeham}
 R^\dag\,H_{\rm diag}\,R &=&  \alpha_0{\mathbf 1} + \alpha_1 \vec{s}_j \cdot \vec{s}_{j+1} \nonumber\\
&=& \alpha_0\, {\mathbf 1} + \frac{\alpha_1}{2}\,\left( (\vec{s}_j + \vec{s}_{j+1})^2 - \frac{3}{2} \right) .
\end{eqnarray}
Since $(\vec{s}_j + \vec{s}_{j+1} )^2$
equals $0$ for a spin-$0$ state and $2$ for a spin-$1$ state, it follows
\begin{eqnarray}
\epsilon_0 &=& \alpha_0 - \frac{3\alpha_1}{4} \nonumber\\
\epsilon_1 &=& \alpha_0 + \frac{\alpha_1}{4}
\end{eqnarray}
Solving for $\alpha_0$ and $\alpha_1$ in terms of $\epsilon_0$ and $\epsilon_1$
\begin{eqnarray}
\alpha_0 &=& \frac{3 \epsilon_1 + \epsilon_0}{4} \nonumber \\
\alpha_1 &=& \epsilon_1 - \epsilon_0  .
\end{eqnarray}
A straightforward computation of the energies of
the lowest spin-$0$ and spin-$1$ eigenstates of $H_{\rm 6-sites}$ gives
\begin{eqnarray}
            \epsilon_0   &=& -2.493577 \nonumber\\
            \epsilon_1   &=& -2.001995 \nonumber\\
				\alpha_0 &=&  -2.124891  \nonumber \nonumber \\
				\alpha_1 &=&    0.491582
\end{eqnarray}
To obtain $h^{\rm conn}(j,2)$ it is necessary to subtract
$h^{\rm conn}(j,1)$ and $h^{\rm conn}(j+1,1)$ from $R^\dag\,H_{\rm diag}\,R $
as follows  
\begin{eqnarray}
	h^{\rm conn}(j,2) &=& R^\dag\,H_{\rm diag}\,R - h^{\rm conn}(j,1)
	- h^{\rm conn}(j+1,1) \nonumber \\
	&=& (\alpha_0+2) {\mathbf 1} + \alpha_1 \vec{s}_j \cdot \vec{s}_{j+1} .
\end{eqnarray}
Finally, given $h^{\rm conn}(j,2)$, the range-2 renormalized Hamiltonian is
\begin{eqnarray}
\label{hrenorm}
H^{\rm ren} &=& \sum_j \big( h^{\rm conn}(j,1) + h^{\rm conn}(j,2) \big)
\nonumber\\
&=& \sum_j \big( (\alpha_0 + 1)\, {\mathbf 1} + \alpha_1 \,\vec{s}_j \cdot \vec{s}_{j+1} \big)
\nonumber \\
&=& V_{\rm thin}\, (\alpha_0+1)\,{\mathbf 1} +  \alpha_1 \,\sum_j \vec{s}_j \cdot \vec{s}_{j+1}
\end{eqnarray}

For an infinite lattice, the fact that the term $V\, (\alpha_0 + 1)\,{\mathbf 1}$
only contributes a constant to the energy density of all states and plays no dynamical role
means that the energy density of the
thinned lattice is $(\alpha_0 + 1)$ plus $\alpha_1$ times the energy
density of the theory we started with.  As in the discussion of the naive
renormalization group, since each site of the thinned lattice
corresponds to three sites on the original lattice we have, according
to the simple range-2 renormalization group approximation, that the energy
density of the spin-$1/2$ HAF, ${\cal E}$, satisfies the following
equation
\begin{equation}
\label{recursone}
	{\cal E} =  \frac{(\alpha_0 + 1)}{3} + \frac{\alpha_1}{3} \,{\cal E} .
\end{equation}
or
\begin{equation}
	{\cal E} =  \frac{\alpha_0 +1}{3\,\big(1-\alpha_1/3\big)} ;
\end{equation}
which is what we obtained by summing the geometric series in our earlier discussion.
Substituting the values of $\alpha_0$ and $\alpha_1$ obtained from the
two-block computation we find
${\cal E}_{\rm ren.grp.}  = -0.4484$ ,
which is to be compared to the exact result ${\cal E}_{\rm exact} = -0.4431$.   
The error in this CORE result, obtained from an exceptionally simple first
principles calculation, is a factor of ten better than that obtained in the naive
renormalization group calculation and a factor of two better than that
obtained from the leading term in Anderson's\cite{Anderson} spin-wave approximation which
assumes that the spin $s$ is a large number and then continues
the answer to $s=1/2$.  Thus, despite the folklore about the difficulty
in improving a real space renormalization group computation, even the simplest
two-state CORE computation, which is only slightly more difficult to carry out than a naive
two state renormalization group computation, produces significant improvements in accuracy.

Since the CORE equation says that the mass-gap of the renormalized theory should
be the same as that of the original theory, the fact that $\alpha_1 < 1$ means that
this gap must vanish. Specifically, since 
$(\alpha_0+1)\,{\mathbf 1}$ plays no role in the dynamics of the
renormalized theory the gap is determined by the range-2 term which is just
$\alpha_1\,\sum \vec{s}_j \cdot \vec{s}_{j+1}$.  But this is just
$\alpha_1$ times the original Hamiltonian and so it follows that the mass gap of the
theory must satisfy the equation
\begin{equation}
	m = \alpha_1\,m .
\end{equation}
Since $0 < \alpha_1 <1$ this means $m = 0$.

\subsection{Spin-$1/2$ HAF: Two-State and Range-$3$ }

Although the preceding discussion shows that CORE computations are, from the outset,
intrinsically more accurate than corresponding naive real-space renormalization group
computations, it remains to be seen that computing more terms in the cluster
expansion improves the answer.  This section presents the results of
a range-$3$ computation for the spin-$1/2$ HAF.

The explicit procedure for calculating the range-$3$ contribution 
is a straightforward generalization of the one followed for the range-$2$
computation.  The space of retained states is now the $8$-dimensional subspace
of the $512$-dimensional Hilbert space of the three block (or nine-site) problem
obtained by taking the tensor product of the lowest lying spin-$1/2$ representation
in each of the three-site blocks.  Since the computation is $SU(2)$ invariant, these
states group themselves into one spin-$3/2$ and two spin-$1/2$ representation and
the computation of 
\begin{equation}
\label{threeblocklimit}
H_{3-block}(j,j+1,j+2) = \lim_{t \rightarrow \infty}
     [[T(t)^2]]^{-1/2}\, [[T(t) H(j,j+1,j+2) T(t)]] \, [[T(t)^2]]^{-1/2}  ;
\end{equation}
is quite simple to carry out. The resulting three-site Hamiltonian
has to be $SU(2)$ invariant and invariant under reflection about its midpoint,
so it must have the form
\begin{eqnarray}
\label{rangethreeH}
	H_{3-block}(j,j+1,j+2) &=& C_3\, {\mathbf 1}_{j,j+1,j+2}\,  + \,\alpha_3\, (\ \vec{s}(j) \cdot \vec{s}(j+1)
	+ \vec{s}(j+1) \cdot \vec{s}(j+2)\ ) \nonumber\\
	& & + \gamma_3\, \vec{s}(j) \cdot \vec{s}(j+2) \\
\end{eqnarray}  
where ${\mathbf 1}_{j,j+1,j+2}$ stands for the three-site unit matrix.
We saw in the previous section that generically
\begin{eqnarray}
	h^{\rm conn}(j,1) &=& C_1\, {\mathbf 1}_j \nonumber\\
	h^{\rm conn}(j,2) &=& (C_2 - 2 C_1)\, {\mathbf 1}_{j,j+1}
	 + \alpha_2\, \vec{s}(j)\cdot\vec{s}(j+1) , \\
\end{eqnarray}
where $ C_2 $ is the coefficient of the unit matrix the expansion
\begin{equation}
	H_{2-block}(j,j+1) = C_2\, {\mathbf 1}_{j,j+1} + \alpha_2\, \vec{s}(j) \cdot \vec{s}(j+1) .
\end{equation} 
and so it follows that
\begin{eqnarray} 
	h^{conn}(j,3) &=& H^{\rm ren}_{3-block}(j,j+1,j+2) - h^{\rm conn}(j,1)
	- h^{\rm conn}(j+1,1) \nonumber\\
	& & - h^{\rm conn}(j+2,1) - h^{\rm conn}(j,2) - h^{\rm conn}(j+1,2)
	\nonumber \\
	&=& (C_3 - 2 C_2 - 3 C_1 )\, {\mathbf 1}_{j,j+1,j+2}
	+ (\alpha_3 -\alpha_2)\, ( \vec{s}(j) \cdot \vec{s}(j+1) + \vec{s}(j+1) \cdot \vec{s}(j+2) )
	\nonumber\\
	& & + \gamma_3 \vec{s}(j) \cdot \vec{s}(j+2)  .\\
\end{eqnarray}
Given these results, after $n$-steps the range-3 renormalized Hamiltonian
will take the form
\begin{eqnarray}
\label{Hofrho}
	H^{\rm ren}_n &=& \sum_j \left( h^{\rm conn}(j,1) + h^{\rm conn}(j,2)
	+ h^{\rm conn}(j,3) \right) \nonumber\\
	&=& \sum_j {\cal C}_n {\mathbf 1}_j + \lambda_n\, \left(\, \vec{s}(j) \cdot \vec{s}(j+1) +
	            \rho_n\, \vec{s}(j) \cdot \vec{s}(j+2) \right) \\
\end{eqnarray}
where I have chosen to write the interaction terms in the Hamiltonian in terms of an
overall scale factor $\lambda_n$, so that the nearest-neighbor interaction always
has a coefficient of unity, and a next-to-nearest neighbor interaction term
which has coefficient $\rho_n$.  As before, accumulating the c-number term ${\cal C}_n$
and dividing by the appropriate power of $3$ yields the ground-state energy density,
which in the case of the range-$3$ computation is $-0.4476$ as opposed to the range-$2$
value of $-0.4483$ and the exact value of $-0.4431$.  Thus, we see that the range-$3$
computation hasn't made a big improvement, we have gone from an error of $1.2\%$
to $1\%$.  The interesting question to ask at this point is why haven't we done
better and is this an inherent limitation of the CORE method?  The answer, as I will
show, is that it is a peculiarity of the range-$3$ approximation and it is worth discussing
in some detail because it shows the way in which the semi-analytic behavior of the CORE
method allows one to easily understand what is happening and what to expect from
the next order computation.  

To understand why the range-$3$ computation fails to produce a bigger improvement in the
the energy density it is convenient to introduce the notion of the
range-$3$ $\beta$-function.  Clearly the $c$-number term ${\cal C}_n$ and the scale
factor $\lambda_n $ enter into the dynamics of the system in a trivial way.  In fact, all
of the dynamics which distinguishes the range-$3$ approximation from the range-$2$ approximation
is encoded in the relative strength of the nearest-neighbor to next-to-nearest neighbor
terms; i.e. the coefficient $\rho_n$.  To understand how the relative strength of these
two terms changes from iteration to iteration it is convenient to define
a function $\beta(\rho)$ as follows: consider a Hamiltonian of the form given in Eq.~\ref{Hofrho}
with $\lambda_n = 1$ and $\rho_n = \rho$ and perform a single range-$3$ CORE transformation to
obtain a new Hamiltonian with new values for $\lambda$ and $\rho$, call
them $\lambda' $ and $\rho' $, then define
\begin{equation}
	\beta(\rho) = \rho' - \rho
\end{equation}

\begin{figure}
\begin{center}
\leavevmode
\epsfxsize=5in
\epsfbox[80 160 530 760]{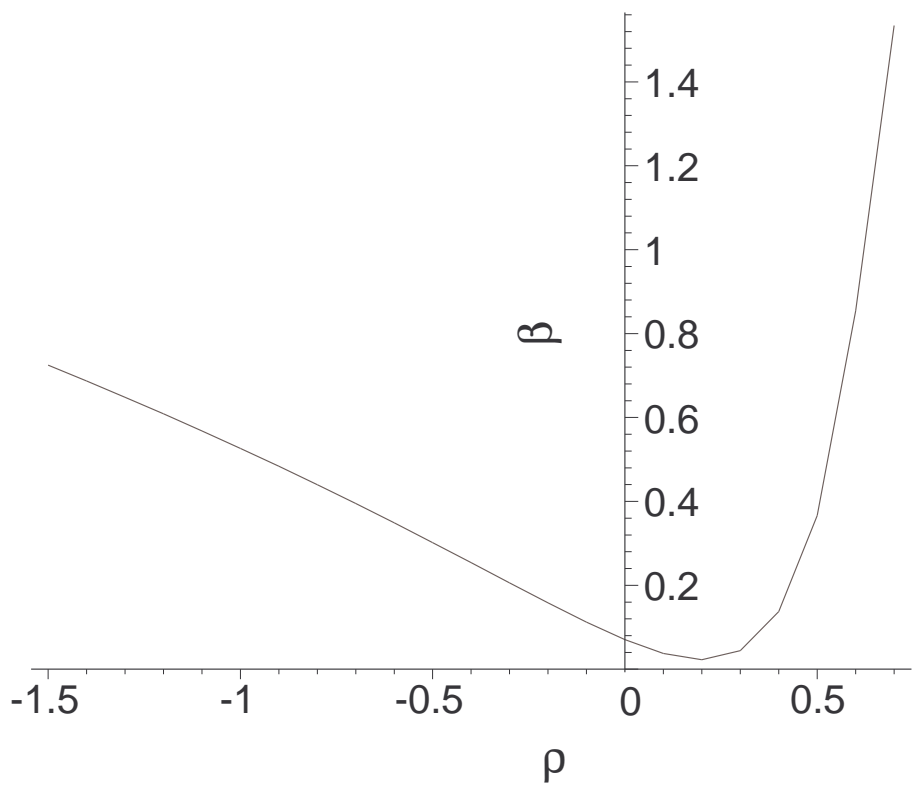}
\end{center}
\caption[betafun]{Range-$3$ CORE $\beta$-function for spin-$1/2$ HAF }
\label{rangethreebetafun}
\end{figure}

A plot of this function for a range-$3$ CORE transformation is show in
Fig.~\ref{rangethreebetafun}.  The starting point for the spin-$1/2$ HAF is the point
$\rho = 0$.  As the figure shows, $\beta(0) > 0$ and so after one transformation
the new theory has a positive value of $\rho $.  Moreover, since $\beta(\rho) > 0$ along the
entire positive axis, we see that with each successive transformation $\rho$ increases without
limit (of course the relevant quantity $\lambda\,\rho$ stays finite) which means that
eventually only the next-to-nearest neighbor term survives and the theory breaks up
into two decoupled HAF's.  This observation tells us immediately what is going wrong
with the range-$3$ computation.  The point is that the range-$3$ computation is done
on a nine-site sublattice and if we ask what happens if we ignore the nearest-neighbor
interaction then we see that theory breaks up into one five-site and one four-site sublattice.
What this means is that even though the infinite volume theory would be two equivalent
decoupled HAF's this part of the computation treats the even and odd sublattices differently.
For example, the groundstate of the five-site sublattice is a spin-$1/2$ multiplet, whereas
the groundstate of the four-site sublattice is spin-$0$.  It is the
asymmetry in the treatment of the two sublattices which causes the spurious growth
of the next-to-nearest neighbor terms relative to the nearest-neighbor term and is
the reason why we don't get the improvement in accuracy that we expected in going
from range-$2$ to range-$3$.  Clearly, if this picture is correct, then doing a range-$4$
computation, which is done on a twelve-site sublattice should correct the problem.  This
is because, if we look at how the range-$4$ computation treats a theory with just
next-to-nearest neighbor couplings, we see that the theory breaks up into two six-site
sublattices and so no asymmetry is introduced into the computation.  The general
features of just this computation is discussed in the next section.

\subsection{Spin-$1/2$ HAF: Two-State and Range-$4$ }

In order to check that our understanding of the origin of the small improvement
in the groundstate energy density for the range-$3$ CORE computation
is correct, it is necessary to
compute the connected range-$4$ contribution to the cluster expansion.  The space
of retained states is now a $16$-dimensional vector space which decomposes into
the sum of one spin-$2$, three spin-$1$ and two spin-$0$ representations of $SU(2)$.
Thus in order to compute the range-$4$ contributions to the CORE formula
we must solve the twelve-site problem and compute the overlap of the retained
states with the eigenstates of the twelve-site Hamiltonian having the appropriate
spins.  The $SU(2)$ symmetry of the problem allows us to treat each sector of
definite total $3$-component of spin separately which still allows us to use
Maple6 to do a brute force computation which takes a reasonable amount of time
on a PC.  As expected, the result for the groundstate energy density in the
range-$4$ computation is  $-0.444286$ which is an error of $-0.28\%$, a better
than three-fold improvement in the error.  Note, the principle change in the
range four computation is not the improvement in the coefficients of the
operators which appear at the range-$3$ level, but is the appearance of a
new set of four-body operators which eventually dominate the renormalization
group flow.  The generic form of the range-$4$ renormalized Hamiltonian is
\begin{eqnarray}
	H^{\rm ren} &=& \sum_j \left[ C_n \,{\mathbf 1}_j + \alpha_1\, \vec{s}(j) \cdot \vec{s}(j+1)
	+ \alpha_2\, \vec{s}(j) \cdot \vec{s}(j+2) + \alpha_3\, \vec{s}(j) \cdot \vec{s}(j+3)
	\right. \nonumber\\
	& &\left. + \beta_1\, \vec{s}(j) \cdot \vec{s}(j+1)\,\vec{s}(j+2) \cdot \vec{s}(j+3)
	+ \beta_2\, \vec{s}(j) \cdot \vec{s}(j+2)\,\vec{s}(j+1) \cdot \vec{s}(j+3) \right.
	\nonumber\\
	& & \left. + \beta_3\, \vec{s}(j) \cdot \vec{s}(j+3)\,\vec{s}(j+1) \cdot \vec{s}(j+2) \right]
\end{eqnarray}     
Table~\ref{rangefouriter} tabulates the energy density and the operator coefficients for the
first eight renormalization group steps.  The two interesting things to note are: first,
the overall scale of all terms in the Hamiltonian drops rapidly and the next-to-nearest
neighbor spin-spin interaction is catching up in strength with the nearest-neighbor
interaction as in the range-$3$ computation; second, the four-body operators become
equal in importance to all of the two-body spin-spin operators. 

\section{CORE and the Spin-$1$ HAF}                        

In the previous sections I focused on the issue of numerical accuracy.  I
showed that simple CORE computations based upon keeping a small number
of states per blcok can, through the cluster expansion, produce very accurate
results.  The next issue which must be addressed is whether simple CORE computations
can provide a better qualitative picture of the physics taking place in a non-trivial
theory.  To show that this is in fact true I now turn to a discussion of the spin-$1$
HAF.  I will show that even the simplest range-$2$ CORE computation 
shows that this theory behaves differently than the spin-$1/2$ theory and
gives a result which is in agreement with the famous Haldane conjecture.

\subsection{The Spin-$1$ Case: Two Versus Three Sites}

In distinction to spin-$1/2$ HAF, the spin-$1$ theory admits a non-trivial two-site
truncation procedure; namely, truncate the nine states of the two-site problem 
to the four-dimensional subspace spanned by its spin-$0$ and spin-$1$ multiplets.  
This truncation procedure leads to a renormalized theory which has four instead
of three states per site and so the form of the Hamiltonian changes.  Subsequent truncations,
however, preserve this new form of the Hamiltonian and give rise to RG-flows which
are easy to compute.  Since the two-site truncation is easy to work with it is the one
for which I will carry out a full numerical CORE computation; however, since
this is different from the procedure we followed for the spin-$1/2$ theory,
I will now show that it is unavoidable.  In other words, I will show that unlike the
spin-$1/2$ case, both the two-site and the three-site blocking procedure forces us to
keep {\it both\/} the lowest lying spin-$0$ and spin-$1$ eigenstates after the first
CORE transformation.  

To see why this happens consider the three-site Hamiltonian of the spin-$1$ HAF,
Eq.~\ref{Hthreesites}.  The difference between the spin-$1/2$ and spin-$1$
three-site Hamiltonians is that in the spin-$1$ case there are more allowed values for
$S^2_{\rm TOT}(1,2,3)$ and $S^2_{\rm TOT}(1,3)$.
Direct substitution of these allowed values into Eq.~\ref{Hthreesites} shows that
the lowest lying $SU(2)$ multiplet for the three-site Hamiltonian is the spin-$1$
representation for which $S^2_{\rm TOT}(1,2,3) = 2$ and $S^2_{\rm TOT}(1,3) = 6$.
Following the dictum of keeping the lowest lying irreducible representation of $SU(2)$
we obtain a renormalized lattice theory which has the same
spin content per site as in the original theory, paralleling the spin-$1/2$
calculation.  The important difference however, is that although the number of states
per site remains the same the range-2 renormalized Hamiltonian takes the more
general form
\begin{equation}
\label{spinonegenform}
   H^{\rm ren} =  \sum_j C\,{\mathbf 1} + \alpha\,\vec{s}(j)\cdot\vec{s}(j+1)
	-\beta\,(\vec{s}(j)\cdot\vec{s}(j+1))^2 . 
\end{equation}

To derive this general form I observe that,
as in the spin-$1/2$ case, the range-1 connected Hamiltonian must be a multiple of the
unit matrix,  since we keep only a single representation of $SU(2)$ per site.
As before, this means that the first non-trivial contribution to the renormalized
Hamiltonian comes from the range-2 terms.  The first contribution
to the connected range-2 Hamiltonian comes from consideration of the two-block
(or six-site) problem.  Since the truncation retains one spin-$1$ multiplet per
block, the retained states of the two-block problem (obtained by taking the
tensor product of the retained spin-$1$ states for each block) span one
spin-$0$, one spin-$1$ and one spin-$2$ representation of $SU(2)$.  The general CORE
rules tell us that the renormalized range-2 Hamiltonian will have these states
as eigenstates, with eigenvalues $\epsilon_0$, $\epsilon_1$ and $\epsilon_2$
(where these stand for the energies of the lowest lying spin-$0$, spin-$1$ and spin-$2$
states of the six-site problem).  One can use a brute force approach
to construct the transformation $R$ and use it to derive the general form
of the connected range-2 term in the original tensor product basis but,
by using a little ingenuity, one can avoid this step.

To carry out the simpler analysis
construct the projection operators $P_0(i,i+1)$, $P_1(i,i+1)$
and $P_2(i,i+1)$ for each pair of sites $i$ and $i+1$ of the renormalized theory; i.e.,
\begin{eqnarray}
\label{projops}
	P_0(i,i+1) &=& \frac{1}{12}\,\left( S^2_{\rm TOT}(i,i+1) - 2 \right)
	\,\left(S^2_{\rm TOT}(i,i+1) - 6 \right) \cr
	P_1(i,i+1) &=& -\frac{1}{8}\,S^2_{\rm TOT}(i,i+1)
	\,\left(S^2_{\rm TOT}(i,i+1) - 6 \right) \cr
	P_2(i,i+1) &=& \frac{1}{24}\,S^2_{\rm TOT}(i,i+1)
	\,\left(S^2_{\rm TOT}(i,i+1) - 2 \right) 
\end{eqnarray}
where the operators $\vec{s}_i$ denote the spin operators
acting on the retained states of the renormalized theory for site $i$
and where I have defined
\begin{equation}
	S^2_{TOT}(i,i+1) = \left( \vec{s}_i + \vec{s}_{i+1} \right)^2 =
	2\,\vec{s}_i \cdot \vec{s}_{i+1} + 4 .
\end{equation}
Without actually computing anything we can now write 
\begin{equation}
\lim_{t\rightarrow\infty} [[T(t)\,H\,T(t)]] = R^{\dag}\,H_{diag}\,R^{\dag}
= \epsilon_0\,P_0 + \epsilon_1\,P_1 + \epsilon_2\,P_2
\end{equation}
which, using Eq.~\ref{projops}, can be immediately rewritten in the form 
given in Eq.~\ref{spinonegenform}.

Now, in order to carry out the next renormalization group step, it is necessary to
reexamine the eigenvalue problem
(for either two or three-site blocks) for generic values of $C$, $\alpha$ and $\beta$.
Of course, since the only important question from the point of view of a CORE
computation is the ordering of eigenstates in the two or three block problem we can,
without loss of generality, set $C = 0$ and  $\alpha = 1$.  Thus, as advertised
in the overview, we see that in order to study the generic problem it is necessary
to start from the Hamiltonian 
\begin{equation}
\label{spinonegenformtwo}
   H^{\rm ren} =  \sum_j \vec{s}(j)\cdot\vec{s}(j+1)
	-\beta\,(\vec{s}(j)\cdot\vec{s}(j+1))^2 . 
\end{equation}
(Note, the value $\beta = 0$ corresponds to the original spin-$1$ HAF.)

\begin{figure}
\begin{center}
\leavevmode
\epsfxsize=5in
\epsfbox[80 160 530 760]{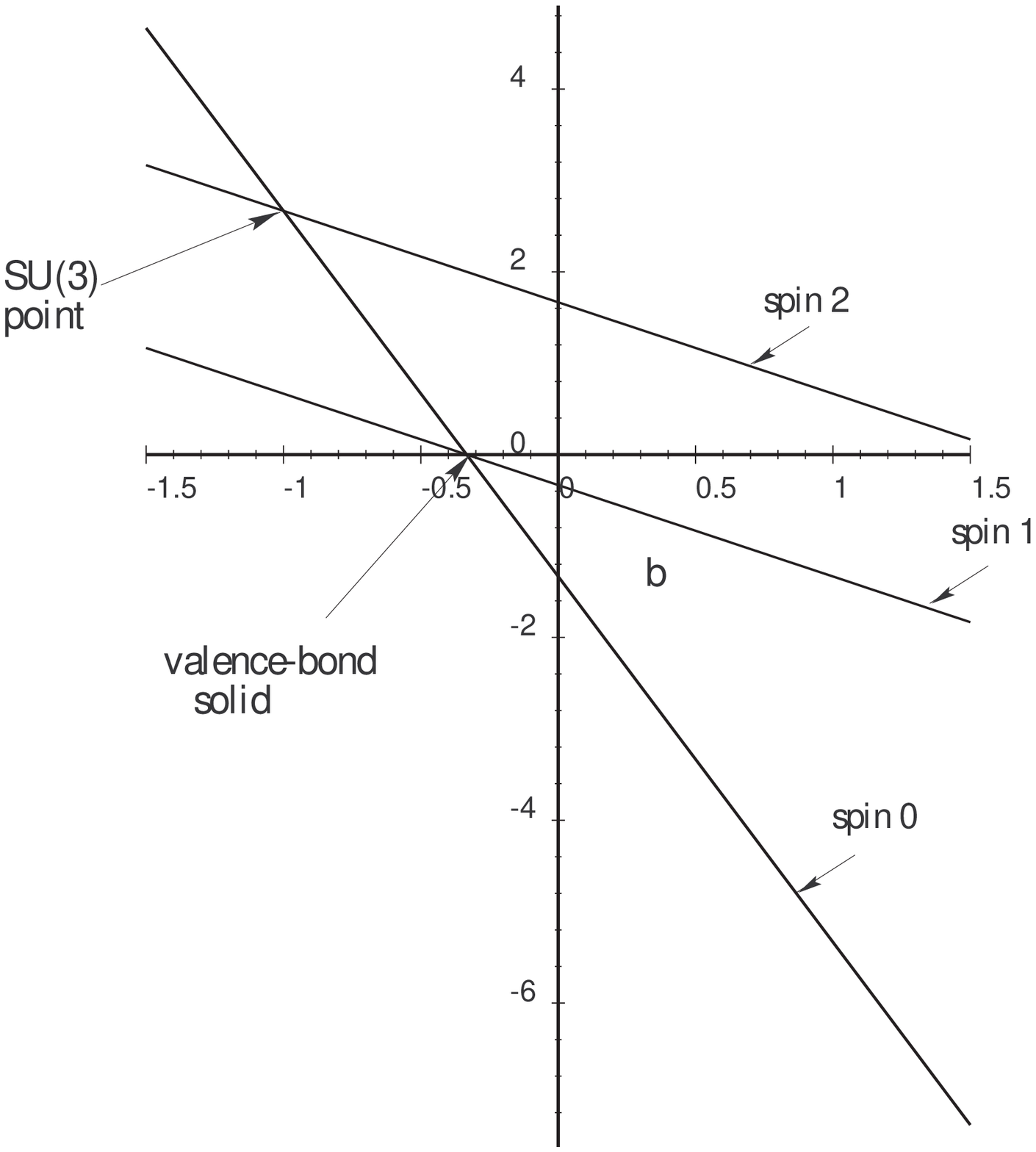}
\end{center}
\caption[ediff]{Energy levels for a two-site block for the Hamiltonian
given by Eq.~\ref{spinonegenformtwo} for $-1 \le \beta \le 1 $.  }
\label{twositemasses}
\end{figure}

\begin{figure}
 \begin{center}
\leavevmode
\epsfxsize=5in\epsfbox[80 160 530 760]{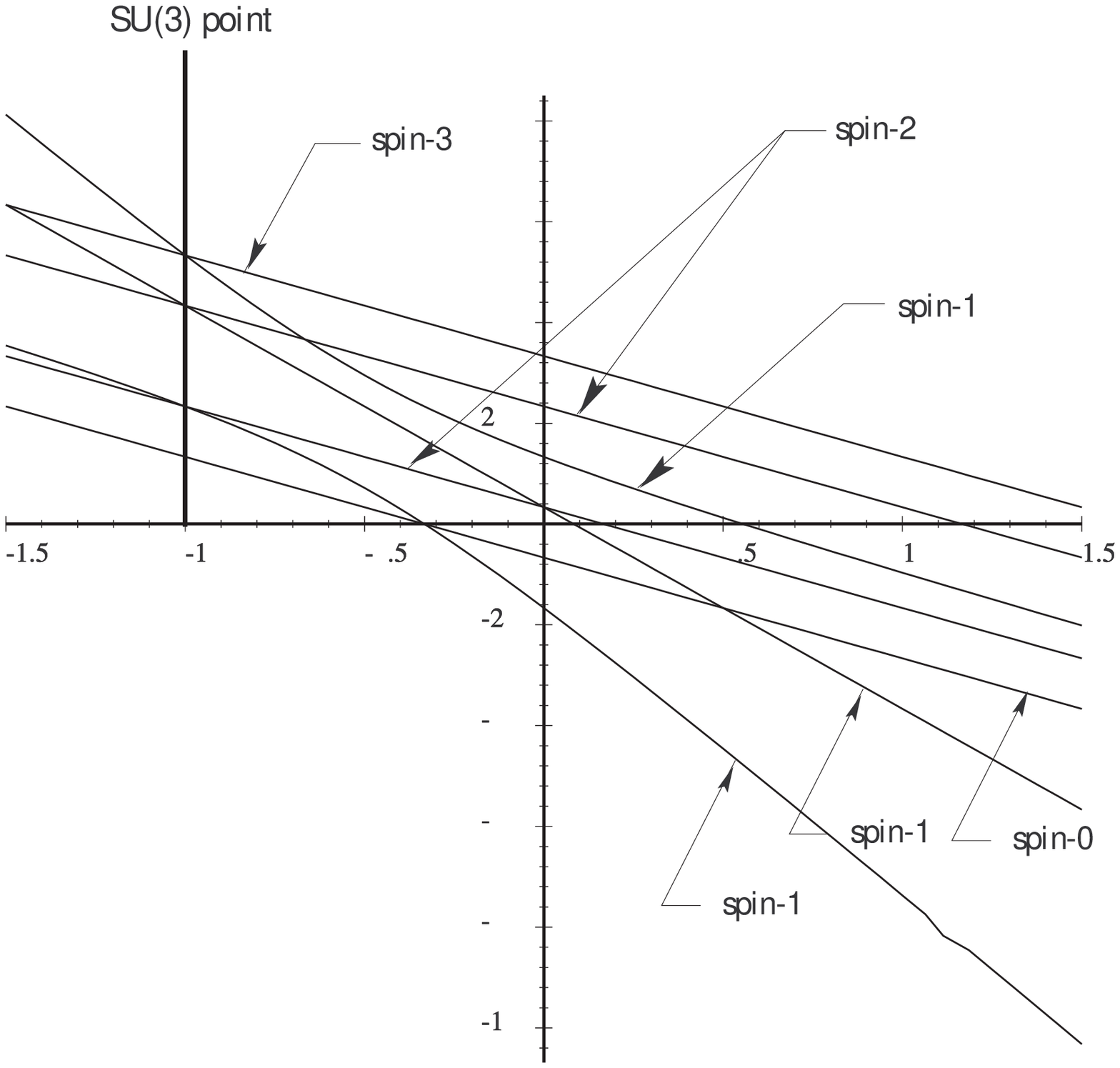}
\end{center}
\caption[ediff]{Energy levels for a three-site block for the Hamiltonian
given by Eq.~\ref{spinonegenformtwo} for $-1 \le \beta \le 1 $.}
\label{threesitemasses}
\end{figure}

The result of diagonalizing the two-site version of this Hamiltonian
for $-1 \le \beta \le 1$ is shown in Fig.~\ref{twositemasses} and the results for the
three-site problem in Fig.~\ref{threesitemasses}, where I have limited discussion
to the range $ -1 \le \beta \le 1 $ for reasons which will become apparent.  
Note that due to the different numbers of eigenstates, etc., these plots look
quite different from one another, however they share several important
common features.  First, observe that the lowest lying spin-$0$ and spin-$1$ state
become degenerate at $\beta = -1/3$ and then cross one another.
This level crossing means, as I said earlier, that any CORE computation
which wishes to treat the region from $-1 \le \beta \le 1$ must keep both multiplets;
i.e., in either the two or three-site case, after the initial renormalization
group step we arrive at a generalized Hamiltonian which forces
us to adopt the two-site prescription of keeping the lowest lying spin-$0$ and
spin-$1$ states.  Second, it is worth noting that something very special
happens at the point $\beta = -1$.  In the two-site case we see that at
this point the lowest lying multiplet is the three-dimensional spin-$1$
representation of $SU(2)$ and that the spin-$0$ and spin-$2$ states become
degenerate and form a single six-dimensional subspace which in fact coincides
with the six-dimensional representation of $SU(3)$.  The degeneracy patterns
shown here demonstrate that the Hamiltonian for $\beta = -1$ can be rewritten
as
\begin{equation}
	H_{\beta=-1} = \vec{Q}(i)\cdot\vec{Q}(i+1)
	\label{SUthree}
\end{equation}
where the $\vec{Q}_i$'s stand for the generators of $SU(3)$.  In this
picture we see that the spin-$1$ representation can be identified as the
triplet representation of $SU(3)$ and the degenerate multiplets of the
two-site problem can be understood to be the $\bar{3}$ and $6$ representations
of $SU(3)$ obtained from the tensor product of two $3$'s.  A brief look at
Fig.~\ref{threesitemasses} supports this picture.  Here we see that at $\beta = -1$
the $27$ states become one one-dimensional multiplet, two eight-dimensional 
multiplets and one ten-dimensional multiplet of degenerate states.  This is,
of course, completely consistent with what would be obtained from the product
of three fundamental triplet representations of $SU(3)$ with the Hamiltonian
given in Eq.~\ref{SUthree}.  This explains my earlier statement that
something interesting happens for $\beta = -1$ and shows that if one really
wished to properly handle this point, it would be necessary to either adopt
a truncation procedure which keeps more states, or one which
goes beyond the range-2 cluster contribution in order to make up for the
violence one is doing to the $SU(3)$ symmetry of the problem.  Clearly,
treating the full $SU(3)$ symmetry of the problem correctly
would require us to eschew a
two-site blocking procedure, since in this case the only non-trivial truncation would
be to a single state.  If we adopted a three-site blocking procedure
then we could adopt a non-trivial truncation based upon keeping nine states,
i.e., the lowest lying singlet and octet representations.  Discussion of
this problem goes beyond the scope of this paper. However I mention it to
explain why one expects from the outset to have trouble using the
four-state truncation algorithm which I will discuss for
values of $\beta \le -1$.

\subsection{Spin-1 HAF: The Calculation}

Since I just finished arguing that generically, after a single
renormalization group step, one will have to deal with a
Hamiltonian of the form
\begin{equation}
\label{betaham}
H = \sum_i \vec{s}(i)\cdot\vec{s}(i+1) - \beta\,(\vec{s}(i) \cdot \vec{s}(i+1))^2
\end{equation}
I will describe the two-block CORE procedure for this generalized
spin-$1$ HAF.  As I already indicated, since this Hamiltonian doesn't have
a single-site term, the first step of the CORE computation is to solve
the two-site problem exactly and truncate to the lowest spin-$0$ and spin-$1$
multiplets of the resulting nine state system (i.e., throw away the spin-$2$
multiplet).  With this choice of projection operator the renormalized
range-1 Hamiltonian is a diagonal $4\times 4$ matrix of the general form
\begin{equation}
	h^{\rm conn}(j,1) = H_{\rm diag} = \left(
	\begin{array}{c c c c}
	\epsilon_0(\beta) & 0 & 0 & 0 \\
	0 & \epsilon_1(\beta) & 0 & 0 \\
	0 & 0 & \epsilon_1(\beta) & 0 \\
        0 & 0 & 0 & \epsilon_1(\beta)
	\end{array} \right)
\end{equation}

To obtain the range-2 term of the renormalized Hamiltonian we have to solve
the two-block or four-site Hamiltonian exactly and use the information
about the exact eigenvalues and eigenstates to construct $R$ and $H_{\rm diag}$.
While in principle $R$ is a $16 \times 16$ matrix, in practice, as in the case
of the spin-$1/2$ HAF, the $SU(2)$ symmetry of the problem greatly simplifies the
job of finding $R$ even though there aren't enough symmetries to render the problem
trivial.  More precisely, the single-block states fall into a spin-$0$ and
spin-$1$ representation of $SU(2)$ so, taking tensor products, we see that the
retained states for the two-block problem are two spin-$0$ representations, three spin-$1$
representations and one spin-$2$ representation of this group.  Clearly, if we
expand any one of the spin-$2$ states in eigenstates of the four-site problem
only states with the same quantum numbers can appear.  Hence, since each of the
spin-$2$ states is distinguished by its third component of spin, each of the
spin-$2$ states will contract onto a different eigenstate of the two-block or
four-site problem but they will all have the same energy.  This argument shows that
the transformation $R_1$ which takes us from the original tensor product basis
to the spin basis is all one has to do for the spin-$2$ states.  Since there are
two independent spin-$0$ representations contained in the tensor product of
the single-block states we have to do a bit more work to fully construct $R$.
To understand exactly what has to be done, let $\ket{\Psi_1}$ and $\ket{\Psi_2}$
denote the spin-$0$ states which can be formed from the $0 \otimes 0$ and $ 1 \otimes 1 $
representations of $SU(2)$.  These states can be expanded in terms of spin-$0$
eigenstates of the two-block problem as 
\begin{eqnarray}
\ket{\Psi_0} &=& a_0 \ket{\phi_0} + a_1 \ket{\phi_1} + a_2 \ket{\phi_2} + \ldots \nonumber\\
\ket{\Psi_1} &=& b_0 \ket{\phi_0} + b_1 \ket{\phi_1} + b_2 \ket{\phi_2} + \nonumber\ldots\\
\label{twostateexp} 
\end{eqnarray}
If, as will generally be the case, both $a_0$ and $b_0$ are non-vanishing, then both
states will contract onto $\ket{\phi_0}$.  One can always avoid this however
by defining rotated states as follows
\begin{eqnarray}
\ket{\chi_0} &=& \phantom{-}\cos(\theta)\,\ket{\Psi_0} + \sin(\theta)\,\ket{\Psi_1} \nonumber\\
\ket{\chi_1} &=& -\sin(\theta)\,\ket{\Psi_0} + \cos(\theta)\,\ket{\Psi_1} 
\end{eqnarray}
where $\cos(\theta) = a_0 / \sqrt{ a_0^2 + b_0^2}$ and
$\sin(\theta) = b_0 / \sqrt{ a_0^2 + b_0^2}$.  With this orthogonal change of basis we
have
\begin{eqnarray}
\ket{\chi_0} &=& \alpha_0 \ket{\phi_0} + \alpha_1 \ket{\phi_1} + \alpha_2 \ket{\phi_2} +
\alpha_3 \ket{\phi_3} + \ldots \nonumber\\
\ket{\chi_1} &=& \phantom{ \alpha_0 \ket{\phi_0} +\  } \beta_1 \ket{\phi_1}
+ \beta_2 \ket{\phi_2}
+ \beta_3 \ket{\phi_3} + \ldots\nonumber\\
\end{eqnarray}
With this definition $\ket{\phi_0}$ is the lowest lying eigenstate of the two-block 
Hamiltonian which appears in the expansion of $\ket{\chi_0}$ and $\ket{\phi_1}$ is
the lowest lying eigenstate which appears in the expansion of $\ket{\chi_1}$; hence,
if one applies $e^{-tH}$ to the rotated states one sees that $\ket{\chi_0}$ contracts
onto $\ket{\phi_0}$ and $\ket{\chi_1}$ contracts onto $\ket{\phi_1}$.

The situation is exactly the same for the spin-$1$ states since the spin-$1$ state made
out of $1 \otimes 0 - 0 \otimes 1$ is even under a reflection about
the middle of the two-site block, whereas the spin-$1$ states made out of
$1 \otimes 0 + 0 \otimes 1$ and $ 1\otimes 1$ are odd under the same reflection.
This means that  the expansion of the even spin-$1$ state cannot contain any eigenstates of
the four-site problem in common with the expansion of the two odd spin-$1$ states.  Thus,
only the two odd spin-$1$ states need to be rotated into one another in order to
guarantee that the lowest lying eigenstate appearing in the expansion of each
state is unique, just as in the spin-$0$ case.

With this behind us, in the rotated basis,  $H_{\rm diag}$ is a matrix whose
diagonal entries are the eigenvalues of the lowest-lying eigenstates
which appear in the expansion of the corresponding rotated state.
Thus,
\begin{eqnarray}
	H_{2-block}(j,j+1) &=& R H_{\rm diag} R^{\dag} \nonumber\\
	h^{\rm conn}(j,2) &=& H_{2-block}(j,j+1) - h^{\rm conn}(j,1) - h^{\rm conn}(j+1,1) 
\end{eqnarray}
Finally, given these results we have the renormalized Hamiltonian defined on the
thinner lattice
\begin{equation}
	H^{\rm ren} = \sum_j \left( h^{\rm conn}(j,1) + h^{\rm conn}(j,2) \right)
\end{equation} 

As with all renormalization group algorithms,
one iterates this process until the sequence of renormalized Hamiltonians either
runs to a fixed point, or until one arrives at a situation which can be handled
by perturbation theory.  The generic step of the recursion follows the
pattern just described, except that now the two-site Hamiltonian
is defined to be
\begin{equation}
	H_{2-site}(j,j+1) = h^{\rm conn}(j,1) + h^{\rm conn}(j+1,1) 
				+ h^{\rm conn}(j,2)
\end{equation}
instead of Eq.~\ref{betaham}.  As before one diagonalizes $H_{2-site}(j,j+1)$
and retains the four lowest lying eigenstates which, if one starts out with
$ -1 < \beta < 1 $, will be a spin-$0$ and spin-$1$ representation of $SU(2)$.
From these states one constructs the new diagonal $h^{\rm conn}(j)$.
Next, one constructs the new range-2 interaction
by using these four states to construct the sixteen retained states
for the two-block problem and expands them in terms of a complete set of
eigenstates for the two-block Hamiltonian. From these expansions one
determines $R$ and $H_{\rm diag}$, from which one immediately 
constructs the new $h^{\rm conn}(j,2)$.  
The results of running such iterations for starting values
of $\beta = -1/3$ and $\beta = 2/3$ are shown in Table~\ref{betaminusonethird} and
Table~\ref{betatwothirds} respectively.

The point $\beta=-1/3$ is one of the special points for which the theory
based upon the Hamiltonian, Eq.~\ref{spinonegenformtwo} is exactly solvable, so it
is interesting to see how the sequence of renormalization group transformations
works for this case.  Table~\ref{betaminusonethird} shows the results of the
first and tenth iterations for the case $\beta = -1/2$.  What is tabulated
for each iteration are the eigenvalues and total spins, $S^2 = S(S+1)$, for
the eigenstates of the renormalized two-site Hamiltonian.  
As we see, initially the sixteen states of the two-site problem fall into
irreducible representations of $SU(2)$ and while the states of each representation
have the same energy, the different representations start out having distinct energies.
This changes with increasing iterations until, as we see in the column for iteration
ten, the system acquires a degenerate spin-$0$ and spin-$1$ multiplet and the remaining
twelve states are all degenerate.  This pattern reproduces itself unchanged for
all succeeding iterations.

To understand what is happening in a simple way it is useful
to rewrite this theory as a theory of spin-$1/2$ states.  This can be easily done
since each site of the lattice has both a spin-$0$ and spin-$1$ representation
living on it and the product of two spin-$1/2$ representations contains exactly
one spin-$0$ and one spin-$1$ representation,  If we identify these representations
with the four states per site of the original theory then we see that
the Hilbert states of the original theory can be set in one-to-one correspondence
with the states of a spin-$1/2$ theory on a lattice
with twice as many sites.  If we identify each two-site block, $B(2j,2j+1)$, with a
single point of the original $\beta = -1/3$ theory, then the range-two reflection
invariant Hamiltonian of the original theory must be equivalent to a generic range-four
Hamiltonian of the form
\begin{eqnarray}
	H &=& \sum_j \left[ \alpha {\bf 1} +  
	A \vec{s}(2j)\cdot\vec{s}(2j+1) + B \vec{s}(2j+1)\cdot\vec{s}(2(j+1)) \right. \nonumber\\ 
	\phantom{\sum_j} &+& C \vec{s}(2j)\cdot\vec{s}(2(j+1)+1) 
	+ D \vec{s}(2j)\cdot\vec{s}(2(j+1)) \nonumber\\
	&+& \left. D \vec{s}(2j+1)\cdot\vec{s}(2(j+1)+1)) \right]
\end{eqnarray}
Now, since for the case $\beta = -1/3$ the spin-$0$ and spin-$1$ states are degenerate
it follows that $A=0$, but at the starting level $B$, $C$ and $D$ do not vanish.  
Clearly one could obtain the exact values of these coefficients from the
values of the level splittings in the first column of Table~\ref{betaminusonethird}.
The more interesting question is what values do these coefficients flow to
as the number of iterations increase.  Although one could do a brute force
calculation of these results it is clear from the eigenvalues appearing in
column two of Table~\ref{betaminusonethird} that the answer is that in this limit
$ A = C = D = 0 $ and $ B = .8359471\ldots $ and $\alpha = 3B/4 $.  With this
choice of parameters we see that of the four spin-$1/2$ sites corresponding
to the two-site block of the original theory, only the inner two spins are
coupled to one another: i.e., the Hamiltonian for the block is just 
\begin{equation}
	H = 3B/4 {\bf 1} + B \vec{s}(2j+1)\cdot\vec{s}(2(j+1)) = B/4 + B (S^{\rm tot}(2j+1,2(j+1))/2-3/4)
\end{equation}
From this we see that if the two inner spins are coupled to a spin-$0$ state
then the two outer spins can be in any configuration (in particular either spin-$0$ or
spin-$1$) producing four states of zero energy, which is what is seen.  Furthermore,
if the two inner spins are coupled to spin-$1$ then one gets $4\times 3=12$ degenerate
states with energy $B$, which is also what is seen.  Turning to the full renormalized
Hamiltonian on the infinite lattice we see that the Hamiltonian describes a fully
dimerized spin-$1/2$ system in which there is no coupling between two spins in the same
block and the block-block couplings only exist between adjacent spins.  It
follows that the ground state of the infinite volume theory is one in which each
pair of neighboring spins is coupled to spin-$0$. Note that this is reminiscent of the
exact solution of this model as a valence bond solid~\cite{magnetbook}.
The lowest lying excited states are those for which any one pair of interacting
spins couples to a spin-$1$ state and all the others couple to a spin-$0$ state.
If one is not at the renormalization group fixed point where $ A = C = D = 0 $,
but a small distance away, where these couplings are small but non-vanishing,
then these degenerate states split into momentum bands.  The interpretation of the
fixed point gap is just the gap to all of the states which have arbitrarily small
momentum in the infinite volume theory.

\begin{figure}
 \begin{center}
\leavevmode
\epsfxsize=5in\epsfbox[80 160 530 760]{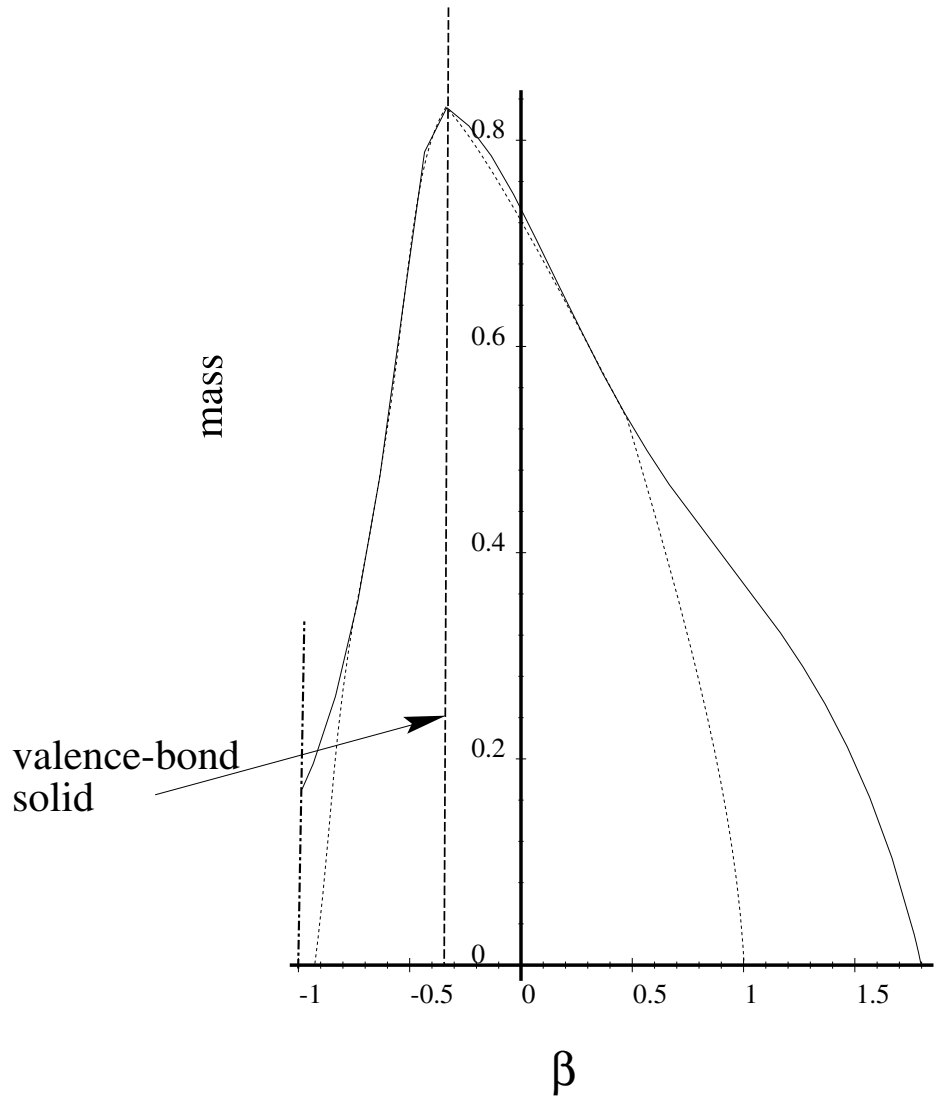}
\end{center}
\caption[ediff]{CORE predicted mass gap for $-1 \le \beta \le 1 $.}
\label{haldanegap}
\end{figure}

If we consider Table~\ref{betatwothirds} we see quite a different picture, in that now
the various multiplets are non-degenerate in the first iteration.
Nevertheless, we see that after ten iterations
the energy eigenvalues (to the accuracy shown) reproduce the
same fixed point pattern as seen in the case $\beta = -1/3$ up to an overall constant.
The only important difference between the case $\beta=-1/3$ and $\beta=2/3$
is that the gap for $\beta = 2/3$ is smaller.
Fig.~\ref{haldanegap} shows the result of carrying out
renormalization group transformations for $ -1 < \beta < 1.8 $.  Thus, the
general picture emerging from this computation is that the spin-$1$ HAF in the region
between $ -1 < \beta < 1 $ is controlled by the valence bond solid fixed point
at $\beta = -1/3$ as one moves away from this point the mass goes down and
at some point both above and below $\beta = 1/3$ the theory appears to become massless.
Given the limitation of the CORE computation to range two terms in the renormalized
Hamiltonian it is not surprising the location of the points where the theory
actually becomes massless is not very accurate.  The dashed curve in Fig.~\ref{haldanegap}
is not meant to be taken seriously, it is drawn in to guide the eye and
remind the reader that the points $\beta = \pm 1$ are
known to be massless theories; one expects that a computation going out to terms
of range three or four would come closer to this picture.  In any event, it seems clear
from the picture that the point $\beta = 0$, which is the spin-$1$ HAF, lies close enough
to the $\beta = -1/3$ theory that one can be confident that it corresponds to a
massive theory as conjectured.  This of course is what we set out to show.

A final point worth commenting upon is the fact that no CORE computations were
done for $\beta \le -1$.  The reason for this is that the truncation scheme used
was to keep only the lowest lying spin-$0$ and spin-$1$ states.
One trouble with this is that the program I used to compute the CORE transformation
simply selected the four lowest lying states, which for the
nondegenerate system in which the spin-$1$ and spin-$2$ have different energies
worked fine.  Unfortunately, this scheme
breaks down at $\beta$ too near $-1$ and one ends up selecting
four states but not necessarily all from either the spin-$0$ or the spin-$1$ multiplet.
In this case one gets spurious results.  To do the
full job correctly would have required a more carefully written program.  
Another problem which contributes to the lack of accuracy of the range-2 calculation
in the vicinity of $\beta = -1$ is that the theory develops an $SU(3)$ symmetry at
$\beta = -1$ and so a truncation scheme which keeps only the spin-$0$ and spin-$1$
multiplets isn't capable of manifestly preserving this symmetry.
A scheme which did preserve the symmetry would need to keep full $SU(3)$ multiplets; i.e.,
the $SU(3)$ singlet state, which corresponds to the spin-$0$ state, and the full
$SU(3)$ octet state, which corresponds to the sum of the spin-$1$ and spin-$2$ states.
Note that while CORE allows one to choose a truncation scheme which doesn't manifestly
preserve the symmetries of the original theory and still obtain correct results,
it does this at the expense of needing longer range couplings in the
renormalized Hamiltonian in order to obtain high accuracy.  

\subsection{General $S$}

In the preceding section I discussed the application of CORE
to the spin-$1/2$ and spin-$1$ HAF, where simple
range-2 arguments sufficed to show that, in agreement with the Haldane conjecture,
the spin-$1/2$ HAF is a massless theory and that the spin-$1$ HAF is massive.
What I did not discuss is what this analysis has to say about the case of
the spin-$S$ HAF when $S$ is greater than one.
A full analysis of the generic case requires doing a 
range-2 computation for all values of $S > 1$, which I have not done.
Nevertheless, examination of the key difference between these
two computations suggests the physics which controls the general case.

To begin the discussion of the HAF for generic $S$
consider the first CORE transformation for an arbitrary $S$ HAF
when one uses a three-site blocking procedure. (The reason for
using a three-site algorithm is that there is no
two-site blocking procedure which works for the spin-$1/2$ HAF.)  For generic $S$
the three-site HAF Hamiltonian is given by Eq.~\ref{Hthreesites}
and the exact solution is as before, only the values for
$S_{\rm TOT}(1,2,3)^2$ and $S_{\rm TOT}(1,3)^2$ change from case to case.  It follows
immediately that the lowest lying representation for the three-site problem
is always spin $S$ and so, the state structure of the renormalized theory is the same
as in the original theory, but as for the spin-$1$ HAF, the Hamiltonian
changes.  As always, truncating to the lowest lying representation yields a range-1
renormalized Hamiltonian which is simply a multiple
of the unit matrix and so, the only real dynamics comes from computing the range-2
terms.  In general, since the single-site retained states form a spin-$S$ representation,
the two-site retained states decompose into a sum of representations going from
$ S^{'} = 0 \ldots 2S$.  Therefore, the new Hamiltonian can be written as a sum
of terms
\begin{equation}
\label{vbsolidham}
	H = \sum_j \sum_{S^{'}=0}^{2S} \epsilon_{S^{'}} P_{S^{'}}(j, j+1)
\end{equation}
where $P_{S^{'}}(j, j+1)$ is the operator which projects the tensor product states
onto the spin-$S^{'}$ representation and $\epsilon_{S^{'}}$ is the eigenvalue of
the lowest lying spin-$S^{'}$ state appearing in the expansion of the projected
tensor product state in terms of eigenstates of the two-block problem.
Again, following the previous discussion, this Hamiltonian can always be rewritten
as a polynomial in the operators $\vec{s}(j) \cdot \vec{s}(j+1)$.  The important thing
to notice at this point is that for integer $S$ and $\epsilon_{S^{'}} = 0$
for $ S^{'} = 0\ldots S$ and $\epsilon_{S^{'}} > 0$, then the Hamiltonian
is a theory of the form constructed by Affleck, Kennedy, Lieb and Tasaki (AKLT)\cite{AKLT}
in order to exhibit theories having a valence-bond solid ground state.
Thus, in the integer spin case any 
three-site CORE transformation immediately maps the integer spin HAF into a theory
which has a massive valence-bond solid theory nearby.  While it would
take doing a complete computation of the CORE flows for this theory in order to
prove that the spin-$S$ HAF lies in the basin of attraction of this theory, it is
exactly what happened in the spin-$1$ case and it is
not unreasonable to conjecture that this is the case for general $S$.
The situation is quite different for theories with half-integer $S$.
In such cases any three-site renormalization group transformation will map
the theory into a sum of half-integral spin representations of
$SU(2)$ with Hamiltonians of the form given in Eq.~\ref{vbsolidham} and it is a theorem
that an AKLT Hamiltonian for half-integral $S$ can't have a valence-bond solid
ground state.  Generically, this result will coincide with what is found in a CORE
computation, since for a half-integer spin a three-site truncation
will always require that one keeps at least one irreducible representation per site
which will perforce have dimension two or greater and these CORE calculations will
generally iterate in a manner similar to the spin-$1/2$ theory; i.e., they will
predict a massless theory, which is consistent with the Haldane conjecture.
Of course, all this is conjecture and a real CORE calculation is needed for some
higher spin theories in order to see how things really work.

\section{Remarks About Correlation Functions}

At this juncture it is important to emphasize that unlike the naive
real-space renormalization group approach one cannot simply calculate
long-distance behavior of a correlation function in the original theory by 
calculating the same function in the renormalized theory.  This is because
these correlation functions change in the same manner as the Hamiltonian
does and they map into more complicated sums of operators which must be
evaluated by the analogous CORE formula.  An example of this is the computation
of the magnetization in the Ising model discussed in Ref.~\cite{COREpaper}.

\section{Conclusion}

In the preceding sections of this paper I exhibited explicit, first principles,
CORE computations for the spin-$1/2$ and spin-$1$ HAF which showed that
CORE is capable of high accuracy even when one keeps only a few states per
block and a few terms in the cluster expansion.  Moreover, I showed that
even a simple range-2 approximation to a full CORE computation agreed for the spin-$1/2$
and spin-$1$ HAF predicts results in agreement with the predictions of the Haldane conjecture.
I also argued that
these computations suggest an attractive picture of how
things can be expected to work for general $S$.  I believe this set of results
shows: first, that the usual folklore, which asserts that all real-space renormalization
group methods which keep only a few states per block will be inaccurate, is incorrect;
second, that even the simplest CORE computations are more than capable of providing
revealing qualitative features which appear subtle from other points of view. 
These results buttress the hope that CORE can fruitfully be applied to the study
of the complicated spin theories which are obtained from free fermion theories
and theories of fermions interacting with gauge-fields which were obtained in Ref.~\cite{QGAF}.
The last point I would like to make is that these arguments show that although
CORE does eventually depend upon one's ability to do numerical computations,
it has a strong semi-analytic flavor and is inherently different from Monte Carlo computations.
CORE computations allow one to focus on the short distance Hamiltonian physics and
the computation of renormalization group flows allows one to directly extract a physical
picture of what is going on.

\newpage
\begin{table}
\caption{Spin-1/2 HAF:  Exact Energy Density = $-\ln(2)+1/4$ = -0.4431472 }
\label{spinhalftable}
\begin{center}
\begin{tabular}{| c | c | c | c |}
\hline
{\bf Range (sites)} & {\bf Energy Density CORE } & {\bf  Pad\'e [N/M]} & {\bf Energy Density }\\
\hline\hline
1  (2)     &   -0.3750000  &            &           \\
\hline
2  (4)     &   -0.4330127  &       &       \\
\hline
3  (6)     &   -0.4387759  &     [1/1]  & -0.4428182    \\
\hline
4  (8)     &   -0.4406777  &     [1/2]   & -0.4431005         \\
           &                  &     [2/1]   & -0.4431022     \\
\hline         
5  (10)   &   -0.44155130  &     [2/2]   & -0.4431337     \\
\hline
6  (12)     &   -0.44202771  &     [2/3]   & -0.4431412         \\
           &                  &     [3/2]   & -0.4431412     \\
\hline         
\end{tabular}
\end{center}
\end{table}

\newpage
\begin{table}
\caption{Spin-1 HAF : Compare to 16-Site Lanczos -1.40293}
\label{spinonetable}
\begin{center}
\begin{tabular}{| c | c | c | c |}
\hline
{\bf Range (sites)} & {\bf Energy Density CORE } & {\bf  Pad\'e [N/M]} & {\bf Energy Density }\\
\hline\hline
1  (2)     &   -1.0000000  &            &           \\
\hline
2  (4)     &   -1.3228757  &       &       \\
\hline
3  (6)     &   -1.3622618  &     [1/1]  & -1.3908701   \\
\hline
4  (8)     &   -1.3771811  &     [1/2]   & -1.3986541         \\
           &                  &     [2/1]   & -1.3986795     \\   
\hline         
\end{tabular}
\end{center}
\end{table}

\newpage
\begin{table}
\caption{CORE flow for range-$4$ spin-$1/2$ HAF}
\label{rangefouriter}
\begin{center}
\begin{tabular}{| r | c | c | c | c | c | c | c |}
\hline
{\bf Iter}  &  ${\cal E} $  & $\alpha_1$ & $\alpha_2$ & $\alpha_3$ & $\beta_1$
& $\beta_2$ & $\beta_3$   \\ \hline\hline
  0   & $-0.000000$ & $1.000000$ & $0.000000$ & $0.000000$ & $0.000000$ & $0.000000$ & $0.000000$ \\ \hline 
  1   & $-0.378606$ & $0.477486$ & $0.033980$ & $0.013715$ & $0.024739$ & $0.069423$ & $0.050919$\\ \hline 
  2   & $-0.436439$ & $0.179218$ & $0.173143$ & $0.007996$ & $0.016203$ & $-0.040886$ & $0.026206$\\ \hline  
  3   & $-0.443547$ & $0.057900$ & $0.005940$ & $0.003327$ & $0.006554$ & $-0.167517$ & $0.010529$\\ \hline 
  4   & $-0.444307$ & $0.016636$ & $0.001615$ & $0.001103$ & $0.002175$ & $-0.005772$ & $0.003595$\\ \hline  
  5   & $-0.444380$ & $0.004270$ & $0.000351$ & $0.000116$ & $0.000632$ & $-0.001756$ & $0.001090$\\ \hline 
  6   & $-0.444385$ & $0.000955$ & $0.000050$ & $0.000076$ & $0.000163$ & $-0.000477$ & $0.000297$\\ \hline 
  7   & $-0.444386$ & $0.000170$ & $0.000003$ & $0.000016$ & $0.000036$ & $-0.000114$ & $0.000073$\\ \hline 
  8   & $-0.444386$ & $0.000014$ & $-0.000007$ & $0.000003$ & $0.000006$ & $-0.000022$ & $ 0.000015$\\ \hline 
\hline
\end{tabular}
\end{center}
\end{table}
\newpage

\newpage
\begin{table}
\caption{CORE flow for case $\beta = -1/3$}
\label{betaminusonethird}
\begin{center}
\begin{tabular}{| c | c || c | c |}
\hline
\multicolumn{2}{| c ||}{\bf Iteration 1}  & \multicolumn{2}{c | }{\bf Iteration 10} \\ \hline\hline
{\bf Levels}  &  {\bf ${\bf S}^2$}  & {\bf Levels} & {\bf ${\bf S}^2$}  \\ \hline\hline
  0               &      0      & 0               &      0      \\  
  0               &      2      & 0               &      2      \\
  0               &      2      & 0               &      2      \\
  0               &      2      & 0               &      2      \\ \hline
  0.89791173      &      6      & 0.83159471      &      6      \\
  0.89791173      &      6      & 0.83159471      &      6      \\
  0.89791173      &      6      & 0.83159471      &      6      \\
  0.89791173      &      6      & 0.83159471      &      6      \\
  0.89791173      &      6      & 0.83159471      &      6      \\ \hline
  0.94191045      &      2      & 0.83159471      &      2      \\
  0.94191045      &      2      & 0.83159471      &      2      \\
  0.94191045      &      2      & 0.83159471      &      2      \\ \hline
  1.1835034       &      2      & 0.83159471      &      2      \\
  1.1835034       &      2      & 0.83159471      &      2      \\
  1.1835034       &      2      & 0.83159471      &      2      \\ \hline
  1.8944584       &      0      & 0.83159471      &      0      \\
\hline
\end{tabular}
\end{center}
\end{table}
\newpage

\begin{table}
\caption{CORE flow for case $\beta = 2/3$}
\label{betatwothirds}
\begin{center}
\begin{tabular}{| c | c || c | c |}
\hline
\multicolumn{2}{| c ||}{\bf Iteration 1}  & \multicolumn{2}{c | }{\bf Iteration 10} \\ \hline\hline
{\bf Levels}  &  {\bf ${\bf S}^2$}  & {\bf Levels} & {\bf ${\bf S}^2$}  \\ \hline\hline
  -0.75395437       &    0      &   -1.6479538       &      0        \\ \hline  
  1.1561163         &      2      &   -1.6479538       &      2      \\ 
  1.1561163         &      2      &   -1.6479538       &      2      \\
  1.1561163         &      2      &   -1.6479538       &      2      \\ \hline
  2.7471518         &      6      &   -1.1820317       &      6      \\
  2.7471518         &      6      &   -1.1820317       &      6      \\
  2.7471518         &      6      &   -1.1820317       &      6      \\
  2.7471518         &      6      &   -1.1820317       &      6      \\
  2.7471518         &      6      &   -1.1820317       &      6      \\ \hline
  3.520943          &      2      &   -1.1820317       &      2      \\
  3.520943          &      2      &   -1.1820317       &      2      \\
  3.520943          &      2      &   -1.1820317       &      2      \\ \hline
  4.6626764         &      0      &   -1.1820317       &      0      \\ \hline
  5.6297153         &      2      &   -1.1820317       &      2      \\
  5.6297153         &      2      &   -1.1820317       &      2      \\
  5.6297153         &      2      &   -1.1820317       &      2      \\
\hline
\end{tabular}
\end{center}
\end{table}

\newpage
\epsfverbosetrue

\newpage

\epsfverbosetrue

\newpage

\epsfverbosetrue
\newpage

\epsfverbosetrue
\newpage

\end{document}